\UseRawInputEncoding
\documentclass[noshowpacs,amsmath,twocolumn,superscriptaddress,aps,prl]{revtex4-1}
\usepackage{setspace}
\usepackage{amsmath}
\usepackage{bm}
\usepackage{graphicx}
\usepackage[nearskip,margin = 0pt]{subfig}

\usepackage{verbatim}
\usepackage{amsfonts}
\usepackage{amssymb}
\usepackage{textcomp}
\usepackage{mathrsfs}
\usepackage{mathtools}
\usepackage{url}
\usepackage{caption}
\usepackage[colorlinks,linkcolor=black,anchorcolor=blue,citecolor=blue,urlcolor=black]{hyperref}
\usepackage{ragged2e}
\usepackage{float}
\usepackage{natbib}

\begin{document}
\title{Extending the spectrum of fully integrated photonics}

%\title{Fully integrated silicon nitride photonics for short wavelength PICs}

\author{
Minh A. Tran$^{1,*}$,
Chong Zhang$^{1,*}$,
Theodore J. Morin$^{2,*}$,
Lin Chang$^{2\dagger}$,
Sabyasachi Barik$^1$,
Zhiquan Yuan$^3$,
Woonghee Lee$^1$,
Glenn Kim$^1$,
Aditya Malik$^1$,
Zeyu Zhang$^1$,
Joel Guo$^2$,
Heming Wang$^3$,
Boqiang Shen$^3$,
Lue Wu$^3$,
Kerry Vahala$^3$,
John E. Bowers$^2$,
Tin Komljenovic$^1$, 
Hyundai Park$^{1\ddagger}$\\
\vspace{3pt}
$^1$Nexus Photonics, Goleta, California 93117, USA.\\
$^2$Department of Electrical and Computer Engineering, University of California, Santa Barbara, CA 93106, USA.\\
$^3$T. J. Watson Laboratory of Applied Physics, California Institute of Technology, Pasadena, CA, USA.\\
$^*$These authors contributed equally\\
\vspace{3pt}
Corresponding authors: $^\dagger$(linchang@ucsb.edu), $^\ddagger$(hpark@nexusphotonics.com).}

%\begin{abstract}

%\end{abstract}

\date{\today}

\maketitle
\noindent
\textbf{Abstract} \\
\textbf{
Integrated photonics has profoundly impacted a wide range of technologies underpinning modern society. The ability to fabricate a complete optical system on a chip offers unrivalled scalability, weight, cost and power efficiency. Over the last decade, the progression from pure III-V materials platforms to silicon photonics has significantly broadened the scope of integrated photonics by combining integrated lasers with the high-volume, advanced fabrication capabilities of the commercial electronics industry. Yet, despite remarkable manufacturing advantages, reliance on silicon-based waveguides currently limits the spectral window available to photonic integrated circuits (PICs). Here, we present a new generation of integrated photonics by directly uniting III-V materials with silicon nitride (SiN) waveguides on Si wafers. Using this technology, we present the first fully integrated PICs at wavelengths shorter than silicon's bandgap, demonstrating essential photonic building blocks including lasers, photodetectors, modulators and passives, all operating at sub-\textmu m wavelengths. Using this platform, we achieve unprecedented coherence and tunability in an integrated laser at short wavelength. Furthermore, by making use of this higher photon energy, we demonstrate superb high temperature performance and, for the first time, kHz-level fundamental linewidths at elevated temperatures. Given the many potential applications at short wavelengths, the success of this integration strategy unlocks a broad range of new integrated photonics applications.
}

\begin{figure*}[!ht]
\centering
\captionsetup{singlelinecheck=no, justification = RaggedRight}
\includegraphics[width=18cm]{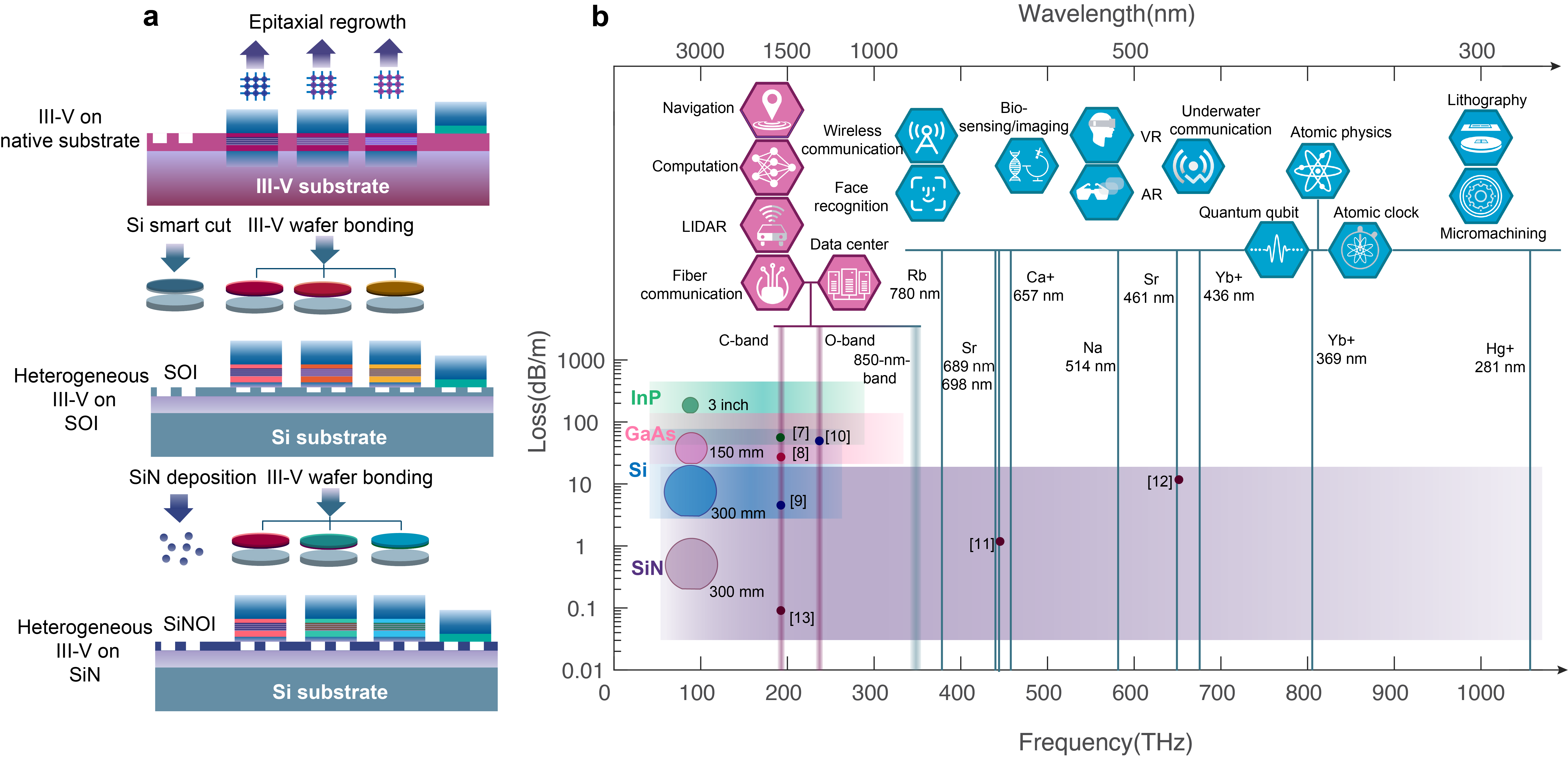}
\caption{\textbf{Fully integrated photonic platforms.} \textbf{a.} The evolution of fully integrated photonic platforms: pure III-V platform relies on multiple epitaxial regrowths to combine active and passives structures; heterogeneous III-V on SOI requires two bonding procedure, the `smart cut' method to produce an integrated Si film and III-V bonding to transfer III-V epi layers from native substrate onto SOI; the heterogeneous III-V on SiN platform needs only SiN direct deposition to integrate SiN film, and only one wafer bonding process to add III-V layer. \textbf{b.} The spectral coverage of fully integrated PICs: boxes represent the transparency window of passive platforms based on different materials (InP \cite{Agostino2015}, GaAs \cite{Ferguson2006}, Si \cite{Biberman2012, Bellegarde2018}, SiN \cite{Chauhan2021,Morin2021,Puckett2021}) that can be used for fully integrated PICs, points represent the current state of the art loss on these passive waveguides, and wafer marker sizes represent the current maximum wafer scale in foundries. The icons on the upper side represent applications of fully integrated PICs over the spectrum map. Purple icons indicate applications accessible to both existing fully integrated PICs and the new III-V/SiN platform; blue icons correspond to new applications made possible by the heterogeneous III-V/SiN platform.
}
\label{fig1}
\end{figure*}

\vspace{3pt}
\noindent
\textbf{Introduction} \\
\\
Integrated photonics has made rapid progress in the last two decades, and the most crucial steps in its advance have been the emergence of novel integration platforms (Fig. \ref{fig1}a). The earliest photonic integration was based purely on III-V materials on native substrates \cite{Merz1985}. Using lattice-matched epitaxial layers, active and passive photonic components were combined on a chip to form optical systems. This approach led to the first generation of commercially viable photonic technologies in communications and fueled the initial push towards sensing and microwave optics applications. Since then, the expansion of the electronics industry has dramatically improved the quality of semiconductor manufacturing, and integrated photonics has benefited from these advances, resulting in high volume application of silicon photonics (SiPh). While III-V manufacturing has not grown apace with silicon, it is possible to manufacture PICs on large-scale silicon on insulator (SOI) wafers by heterogeneously bonding III-V epitaxy in a variety of different ways \cite{Komljenovic2018}. Leveraging mature CMOS foundry infrastructures, the SOI integrated photonics platform significantly reduces the cost of photonic chips at scale. 

Another key factor driving the evolution of integrated photonics is low passive waveguide propagation loss. Since SOI waveguides exhibit propagation losses an order of magnitude lower than III-V waveguides \cite{Margalit2021}, SiPh PICs can accommodate more individual components and thus support more complex photonic systems. Moreover, lower loss boosts the performance of passive structures and coherent light sources. Capitalizing on these advantages, along with reduced cost and large scale production, SiPh has seen explosive growth in the last few years, opening up a plethora of new applications from neural networks \cite{Zhang2021} to Lidar \cite{Poulton2017} to quantum photonics \cite{Qiang2018}.

With the broadening of the application scope, however, the limitations of the SOI platform are beginning to surface. One such limitation comes from the band gap wavelength of silicon, around 1.1 \textmu m (Fig.1b). Below this wavelength SOI waveguides become highly absorptive. Therefore, ultraviolet, visible and a substantial portion of the near infrared are currently inaccessible to state-of-the-art integrated photonics. This restriction is a bottleneck that prohibits on-chip solutions in such fields as atomic physics, augmented reality (AR)/virtual reality (VR), biosensing and quantum communications \cite{Mehta2020, Niffenegger2020, Hummon2018, Masood2021, Zinoviev2011, Orieux2016, Blumenthal2020} as shown in Fig. \ref{fig1}b. 
 
One promising path towards addressing this problem is to implement passive structures using Silicon Nitride (SiN), another CMOS compatible material with a bandgap wavelength below 300 nm \cite{Kruckel2017}. In addition to supporting short wavelengths, SiN waveguides have been shown to exhibit losses $<$ 0.1 dB/m at telecom wavelengths \cite{Jin2021, Puckett2021} and remain scattering-limited down to below 460 nm \cite{Morin2021}, making them attractive for ultra-high $Q$ microcavities, narrow linewidth lasers and integrated nonlinear devices such as microcomb sources and on-chip frequency converters. Furthermore, because SiN wafers are produced by direct deposition on a Si substrate, they do not require any expensive smart-cut process, suggesting an opportunity to further reduce the cost of foundry manufactured PICs, increase scale, and thus further extend the photonics application space. 

However, until recently, the integration of active components onto SiN PICs has been impeded by the large index mismatch between SiN ($\approx$2) and III-V materials ($>$3). SiN and III-V structures have been integrated on the same substrate to form highly coherent laser sources at telecom wavelengths, but such integration has relied on an intermediary Si layer for passive-active transitions, which still prohibits short wavelength operation \cite{Xiang2020, Beeck2020}. 

In the present work, we report a new generation of integrated photonics by directly bridging active and passive elements in a heterogeneously integrated III-V/SiN platform. Within this integration scheme, we demonstrate the first fully integrated photonic platform using submicron wavelengths that provides versatile building blocks including lasers, semiconductor optical amplifiers (SOAs), modulators, photodetectors and various passive elements. The combination of III-V gain section with SiN external cavities yields the first heterogeneously integrated, narrow linewidth, widely tunable laser operating below the band gap wavelength of Si, a device with tremendous implications for atomic physics, sensing and precise metrology. Moreover, the short wavelength platform exhibits superior high temperature performance among coherent light sources, which can be used to improve power-efficiency in datacenters and other hot environments. With numerous potential uses in both academia and industry, this new integrated photonic platform opens the door to many new applications and heralds a new generation of integrated photonics technologies.

\noindent

 \begin{figure*}[!ht]
\centering
\captionsetup{singlelinecheck=no, justification = RaggedRight}
\includegraphics[width=18cm]{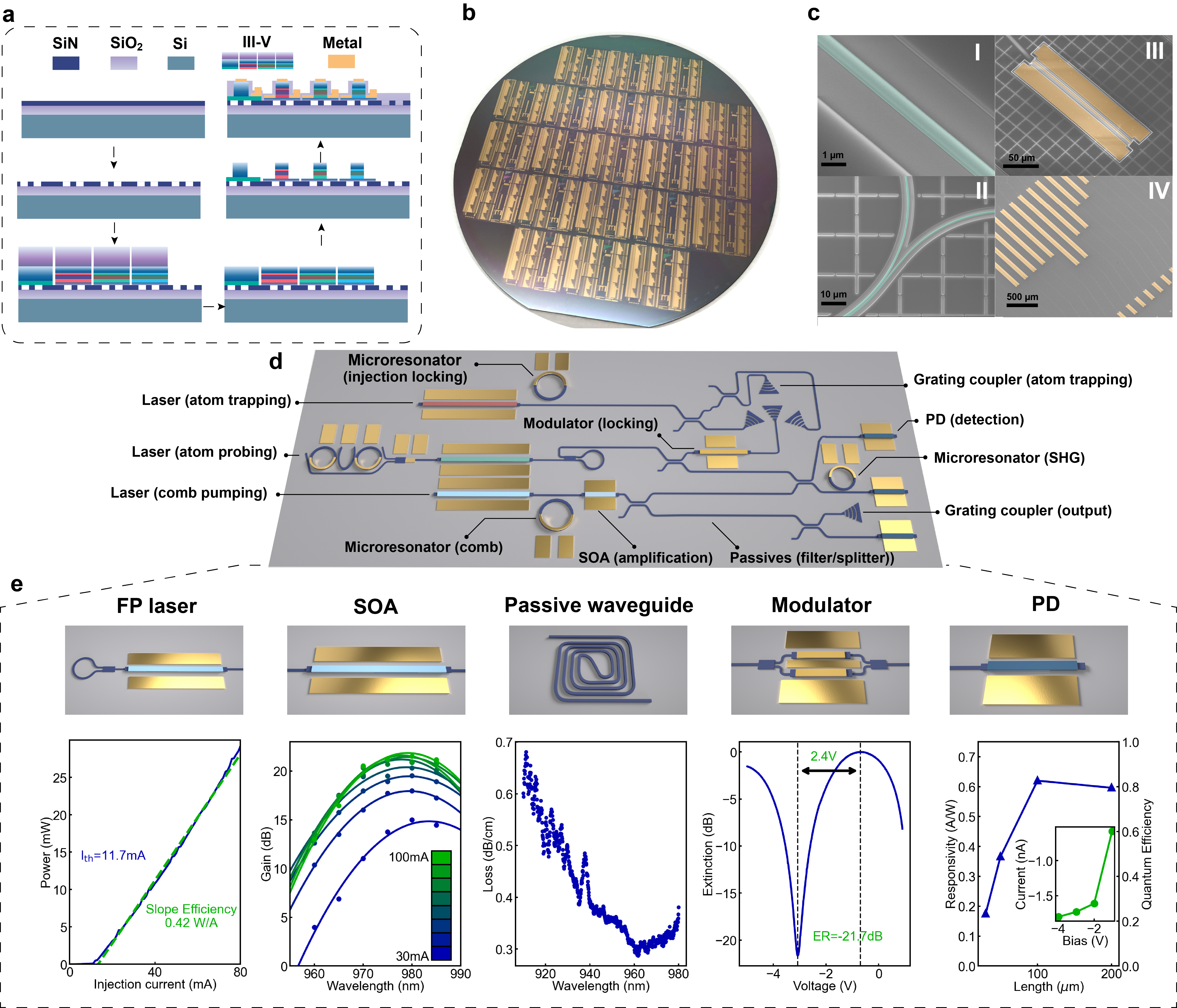}
\caption{\textbf{Silicon nitride heterogeneous photonics platform with a full set of passive and active building block components supporting sub-\textmu m wavelengths.} \textbf{a.} Simplified wafer-scale process flow. Steps shown: (i) SiN deposition on a thermally oxidized Si substrate; (ii) SiN waveguide patterning; (iii) Bonding of multiple III-V epitaxial structures; (iv) Substrate removal of the III-V epitaxy; (v) III-V processing, including multiple dry/wet etches to form the p-n junctions for active devices; (vi) Dielectric cladding deposition, via etches and metallization that complete the device fabrication. \textbf{b.} A photograph of a fully-processed 4-inch wafer that contains thousands of devices . \textbf{c.} SEM images of (i) a SiN waveguide (ii) a waveguide coupler (iii) a III-V waveguide and electrical contacts for active components (iv) An array of lasers and photodiodes connected by SiN waveguides. \textbf{d.} Envisioned schematic of a fully integrated atomic clock system fabricated on a single chip \textbf{e.} Active and passive functionalities supported on the platform, with characteristic performance. \textbf{FP laser.} A Fabry-Perot laser with integrated broadband mirrors that has $<$12 mA current threshold and more than 25 mW output power to a SiN waveguide. \textbf{SOA.} a semiconductor optical amplifier with maximum gain of 22 dB at 980 nm with 100 mA bias current. The 3 dB bandwidth of the gain at 100 mA spans over 20 nm.  \textbf{Passive waveguide.} SiN waveguides with sub dB/cm propagation loss in the 900-980 nm wavelength range. \textbf{Modulator.} A Mach-Zehnder interferometer with phase modulators showing V$_\pi$=2.4 V and $>$20 dB extinction ratio. \textbf{PD.} A  photodiode with $>$0.6 A/W responsivity at 980nm and nA-level dark current.
}
\label{fig2}
\end{figure*}
\noindent
\textbf{Heterogeneously integrated III-V on SiN photonics platform} \\
Heterogeneous III-V/SiN photonic devices consist of III-V based epitaxial layer structures bonded on top of SiN waveguides. A simplified fabrication process flow for the III-V/SiN heterogeneous photonic devices is illustrated in Fig. \ref{fig2}a. A silicon substrate is thermally oxidized to form a suitable SiO$_2$ base layer for waveguide cladding. Next, a SiN film is deposited and patterned with a photolithographic stepper system and dry etched to form passive waveguide structures. III-V epitaxy is directly bonded onto the nitride wafers in an optimized molecular bonding process. The III-V substrates are then removed by mechanical polishing and a selective wet etch before continuing the III-V process to form active components. A blanket dielectric is deposited to form the top cladding for both SiN and III-V waveguides, as well as the insulator between metal sections. Finally, vias are opened and metal pads are deposited to form electrical contacts to the devices. Fig. \ref{fig2}b shows a photograph of  a completed wafer with hundreds of lasers fabricated on a 4-inch silicon substrate. Scanning electron microscope (SEM) images (Fig. \ref{fig2}c (i-iv) show a single SiN waveguide, a coupler, a III-V waveguide with III-V/SiN coupler on one side, and an array of lasers connecting with an array of photodiodes via SiN waveguides, respectively.

An essential feature of the platform is efficient light coupling between III-V and SiN waveguides. Due to the large refractive index of III-V material compared to SiN, the optical mode of a III-V/SiN heterogeneous waveguide is fully localized in the III-V layer. This is a fundamental distinction from a typical III-V/Si heterogeneous waveguide where the similar refractive indices of Si and III-V make it possible for the optical mode to hybridize in both materials \cite{Fang2006}. As a result, the usual adiabatic coupling scheme based on evanescent fields, though well suited to III-V/Si photonics, does not serve well for III-V/SiN. Butt-coupling, a non-adiabatic method widely used in conventional optics, is advantageous in this case. However, efficient butt-coupling requires maximal spatial overlap between the waveguides being coupled, which is not obtainable in a wafer-scale heterogeneous integration platform because the III-V and SiN layers, being bonded together, cannot be vertically aligned. To solve this problem, we have introduced a novel single III-V/SiN coupler structure that make use of both aforementioned coupling schemes. The III-V/SiN coupler is an intermediary waveguide formed by patterning the dielectric cladding in between III-V and SiN waveguides. On one end, the geometry of the intermediary waveguide is optimized for butt-coupling to the III-V waveguide and on the other end, it is optimized for adiabatic evanescent coupling to the SiN waveguide. This process is described in greater detail in \cite{park2020pat}. A coupling efficiency of up to 70\% was demonstrated in the first generation, and 90\% efficiency is achievable with optimal designs \cite{Park2020}. 
 
 This platform, by directly combining III-V with SiN, brings together versatile functionality into a single, fully-integrated, short wavelength PIC ecosystem. Fig. \ref{fig2}d shows a proposed integrated PIC for an integrated atomic clock system which can be realized with the building blocks demonstrated on this platform (see Fig. \ref{fig2}e). We have here characterized devices operating around 980 nm to illustrate these capabilities. All of the essential active photonic components are implemented. Fabry-Perot (FP) lasers, formed with near 100\% loop mirrors on the back side and 10\% mirrors on the front side, provide a basic light source. They exhibit a low threshold current of 12mA while the output power and slope efficiency exceed 25mW and 0.38 W/A, respectively; Integrated semiconductor optical amplifiers (SOAs) with more than 22dB optical gain and 20 nm 3-dB bandwidth are crucial components for loss compensation, as the number of components and complexity in PICs are ever growing. This element is particularly important in applications such as communications and sensing where the signal to noise ratio (SNR) is an essential metric. For photon detection, photodiodes (PDs) exhibit more than 0.6A/W responsivity, 80\% quantum efficiency and nA-level dark current. Having such high-performance PDs directly integrated along with other components is invaluable for quantum applications, since it allows direct readout of optical signals without the I/O coupling loss. We also demonstrate phase shifters  with a V$_\pi$ of only 2.4 V and Mach-Zehnder modulators with more than 22 dB extinction ratio, which can be used to encode information for communications \cite{chen1970}, create side bands for PDH locking \cite{Drever1983} or manipulate atomic states \cite{LaserCoolingbook}, which can be directly driven by CMOS driver. Complementing the III-V active elements are SiN passive waveguides, achieving sub-dB/cm propagation loss, with the lowest reaching below 0.3 dB/cm.  %\ Based on these high $Q$ factor ring-based resonators, we realized spectral filtering functionalities. Similar rings can also be used for microcomb generation using $\xi_3$ nonlinearity \cite{Gaeta2019} or three wave mixing based on the recently discovered photogalvanic effect in SiN \cite{Lu2021}.
 It is also worth noting that our recently developed ultra-low loss SiN waveguides \cite{Morin2021, Jin2021}, which will be employed in this platform in the future, can further reduce waveguide loss by two orders of magnitude.\newline

 \begin{figure*}[!ht]
\centering
\captionsetup{singlelinecheck=no, justification = RaggedRight}
\includegraphics[width=18cm]{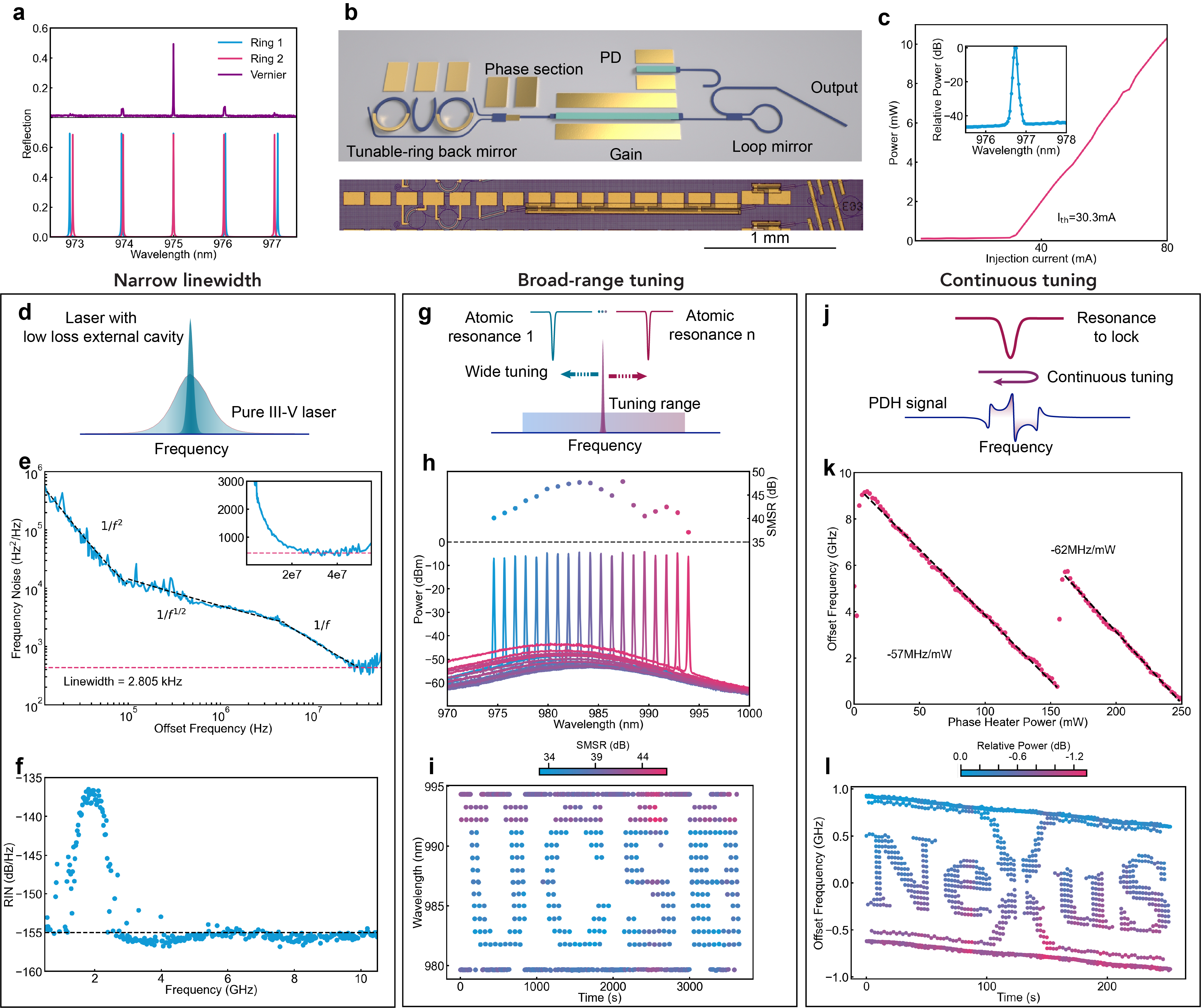}
\caption{\textbf{Integrated coherent, widely tunable lasers on silicon nitride} \textbf{a.} The wavelength response of the individual ring resonators and the resulting measured Vernier spectrum when the two rings are cascaded. The ring resonators' combs are slightly different in free spectral range (FSR) to construct a single resonant line based on Vernier effect. \textbf{b.} Schematic of a dual-ring tunable laser with a back-mirror formed by two ring-resonators placed within a 100\% loop mirror, a front loop mirror with 50\% power reflectivity and a GaAs based SOA section in between. Thermal micro-heaters are fabricated on the rings and a part of laser cavity to align the rings, select the wavelength and tune the round trip phase accumulation. The photograph shows a tunable laser chip with a form factor of $<$3$\times$0.3 mm$^2$. \textbf{c.}  LI characteristic of the laser ata fixed wavelength, showing a threshold current of 30.3 mA and $>$10 mW output power, with single mode lasing spectrum shown in the inset. \textbf{d.} Linewidth and noise reduction achieved with low loss SiN waveguide external cavity. \textbf{e.} Frequency noise measurement shows a fundamental linewidth of 2.8 kHz, a record for integrated lasers below Si bandgap wavelength. \textbf{f.} Lower than -155 dB/Hz level relative intensity noise (RIN) outside the relaxation oscillation resonance. \textbf{g.} Wide tuning range enables access to many atomic resonances \textbf{h.} Broad wavelength tuning enabled by Vernier effect. Wavelengths are tuned across $>$20 nm wavelength tuning range with high side mode suppression ratios (SMSRs) accross the whole range. \textbf{i.}  A "UCSB" logo created by stepping the wavelength of the laser over time. The color of each dot indicates the measured SMSR at that specific time-step \textbf{j.} Mechanism of locking a resonance with a single continuous tuning parameter, crucial for locking to atomic transitions. \textbf{k.} Mode hop free, continuous tuning of the III-V/SiN laser frequency obtained over $>$ 8 GHz by sweeping the phase tuning section alone. \textbf{l.} A "Nexus" logo created by tuning the laser frequency without mode hop, showing great stability and precise control within a 2 GHz frequency range.
}
\label{fig3}
\end{figure*}

\noindent
\textbf{Integrated coherent laser beyond silicon bandgap} \\
\\
With heterogeneous photonics, it is possible to integrate low loss passive waveguides with optically versatile active materials. At the telecom-band, for example, low-loss silicon waveguides have been paired with InP-based optical gain material to produce integrated narrow linewidth lasers \cite{Tran2019}. In the present work, by uniting high quality SiN passives with short-wavelength III-V gain medium, our platform for the first time offers a similar capability beyond the silicon bandgap limit.

An integrated laser operating at 980 nm, which consists of a GaAs gain region and a SiN-external-cavity, is presented as a proof of concept. Fig. \ref{fig3}b shows the schematic design of the laser, whose back mirror consists of two ring resonators cascaded within a loop in an add-drop configuration. Each ring resonator forms a comb in the wavelength domain with adjacent comb-lines separated by one free-spectral-range (FSR). As shown in (Fig. \ref{fig3}a), by choosing radii for the rings so that the FSRs of the two combs are slightly different, the dual-ring mirror’s reflective spectrum - which is the product of these two combs - is a Vernier comb that has only a single dominant comb line  at which the two individual comb lines align. In addition to the Vernier tunable ring mirror, the laser also features a phase tuning element and an on-chip monitor photodiode. The output power from the laser is greater than 10 mW near the gain peak, as shown in the LI curve in Fig. \ref{fig3}c, where the power is measured while wavelength is maintained around 976.5 nm. For a fixed gain current of 75 mA, the power output is measured to be higher than 6 mW across the whole wavelength range.

A scalable, compact III-V/SiN heterogeneous laser, with a footprint of less than 1 mm$^2$ as shown in Fig. \ref{fig3}b, is valuable for a wide range of applications at short wavelength range \cite{Wieman1991}. One important example is atomic physics. At present, precise atomic cooling and probing systems predominantly rely on bulky optical sources such as external cavity diode lasers (ECDLs), which cannot be easily mass-produced \cite{Arnold1998, Littman1981}. The III-V/SiN heterogeneous laser described here offers performance comparable to an ECDL but with the form factor of  a fully integrated device. Fig. \ref{fig3}e shows the double-sided power spectral density (PSD) of the laser noise at 980 nm wavelength, measured with a delayed self-heterodyne setup and cross-correlation technique (see Methods). The spectrum is dominated by 1/f$^x$ noise at low off-set frequency range as commonly observed in semiconductor lasers. At around 30 MHz offset frequency, a white noise floor of ~450 Hz$^2$/Hz is reached, corresponding to a fundamental linewidth of 2.8 kHz, a record low linewidth for fully integrated lasers operating below Si bandgap wavelength. Unlike previous integrated pure III-V lasers, whose fundamental linewidths (typically above 100 kHz \cite{Margalit2021}) are broader than many atomic transition linewidths \cite{Vogel1999, McFerran2010}, the III-V/SiN heterogeneous laser presented here derives significant noise reduction from its low-loss SiN ring resonator-based mirror, opening access to those narrow line atomic transitions. The III-V/SiN heterogeneous laser also shows good amplitude noise performance, with relative intensity noise (RIN) lower than -155 dB/Hz (noise floor of the measurement tool) outside the relaxation oscillation resonance near 2 GHz offset frequency, as shown in  Fig. \ref{fig3}f.

Besides its narrow linewidth and low noise, another key feature of the Vernier laser design is its wide tunability which has not previously been available for short wavelength integrated photonics. With only narrow tuning capability, producing specific wavelengths (e.g. targeting atomic transitions) demands tight fabrication tolerances. Using micro-heaters placed on top of the ring resonators, one can make use of the thermo-optic effect to tune each ring’s comb, shifting of the Vernier location to the desired wavelengths. This simple Vernier comb principle provides a mechanism to obtain a reconfigurable optical filter on chip -- the key to a widely tunable laser. Fig. \ref{fig3}h shows the lasing spectra measured by coarsely stepping the wavelength in ~1 nm increments, characterized at 25$^{\circ}$C. The spectrum at each step is taken after tuning the rings and phase tuner to maximize the photocurrent at the monitor photodiode. The tuning range is about 20 nm (equivalent to approximately 6 THz), primarily limited by the gain bandwidth from the 980 nm quantum wells. The lasing side-mode-suppression-ratio (SMSR) is greater than 35 dB across the entire tuning range and approaches 50 dB when the lasing wavelength is located near the gain peak as shown in the inset. The wavelength of the laser can be stepped repeatably over a wide range without sacrificing SMSR, as shown in Fig. \ref{fig3}i, where the y-axis shows the laser’s lasing wavelength as a function of time and the dot color indicates the SMSR of the lasing mode.

In addition to broad tuning, when locking a laser to a high-Q cavity or atomic transitions, continuous fine tuning is often required over a smaller range. In another emerging application, FMCW Lidar systems, the range of continuous frequency chirping determines the accuracy of detection \cite{uttam_FMCW1985}. As shown in Fig. \ref{fig3}k, by simply sweeping the phase tuner, our laser supports a mode-hop free tuning range of ~8 GHz. Note that a much larger mode-hop free tuning range can be achieved by simultaneously tuning the rings and the phase section \cite{Rees2020}. As shown in Fig. \ref{fig3}l, the frequency of the laser can also be controlled very precisely over several GHz. \newline

\noindent

 \begin{figure*}[!ht]
\centering
\captionsetup{singlelinecheck=no, justification = RaggedRight}
\includegraphics[width=18cm]{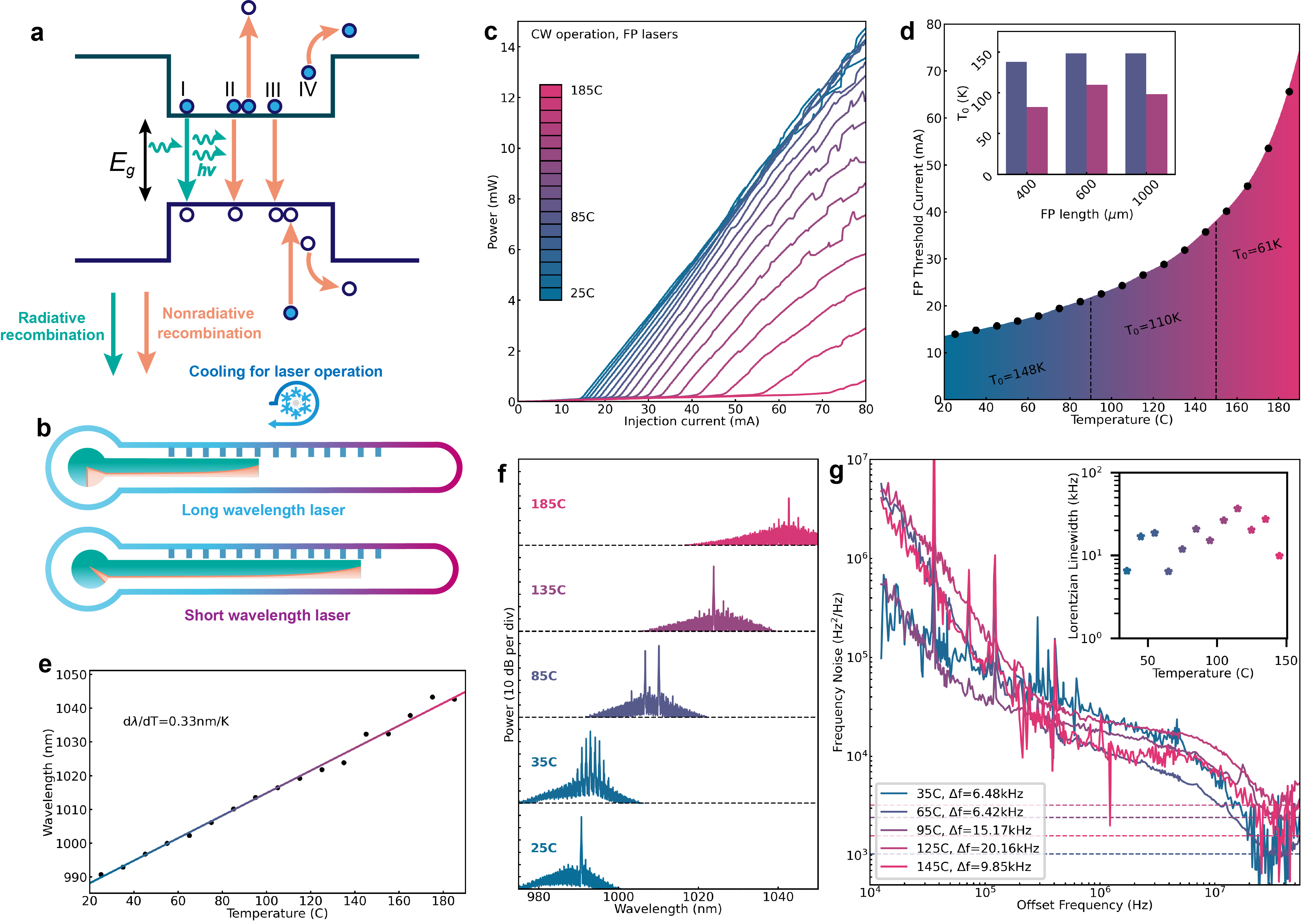}
\caption{\textbf{Extremely high temperature fully integrated lasers.} \textbf{a.} Simplified band diagram of laser and illustration of major carrier recombination and leakage processes, including (I) radiative recombination, (II) Auger recombination, (III) Intervalence band absorption (IV) Carrier leakage over the hetero-barriers. \textbf{b.} The temperature dependence of carrier recombination processes for fully integrated long wavelength lasers and the short wavelength lasers in this work. Nonradiative recombination increases exponentially with temperature, but the effect is reduced in a short wavelength GaAs platform due to the increased energy band gap and quantum well depth. The allowed working temperature is represented by the length of the solid bar, and cooling process is necessary if the free-running temperature of the device goes beyond the working temperature range.  \textbf{c.} LI characteristics of SiN heterogeneous Fabry-Perot lasers from 25$^{\circ}$C to 185$^{\circ}$C. \textbf{d.} Threshold current of the laser vs. temperature, extracted from the LI curves. Characteristic temperature $T_0$ = 148K within 20$^{\circ}$C--90$^{\circ}$C; $T_0$ = 110K in the range from 90$^{\circ}$C to 150$^{\circ}$C, the highest lasing temperature reported in a commercial heterogeneously integrated laser \cite{Jones2019}, and $T_0$ = 61K above 150$^{\circ}$C. Inset: Characteristic temperatures $T_0$ in the first two ranges (20$^{\circ}$C-90$^{\circ}$C and 90$^{\circ}$C--150$^{\circ}$C) of FP lasers with various cavity length. \textbf{e.} FP lasing wavelength vs. temperature, illustrating linear gain red shift of 0.33 nm/K. \textbf{f.} FP lasing spectra at select temperatures.  \textbf{g.} Select frequency noise spectra of the single mode tunable laser measured at elevated temperatures, showing $<$10 kHz fundamental linewidth even at 145$^{\circ}$C. Inset: fundamental linewidth versus temperature from 35$^{\circ}$C to 145$^{\circ}$C.
}
\label{fig4}
\end{figure*}

\noindent
\textbf{High-temperature advantage of short-wavelength PICs}\\ 
\\
A major challenge for integrated photonics is the requirement of active cooling. Since the performance of diode lasers degrades at elevated temperature, it is necessary to cool the PICs to maintain performance, contributing significantly to the power consumption of data centers, which account for more than 1 \% of the total power consumption of humanity \cite{Jones2018}. Thermal degradation of lasers is caused by gain reduction due to the wider spreading of the Fermi distribution of carriers at increased temperature \cite{Piprek2000} and by the loss of radiative carriers via various mechanisms, notably including carrier leakage over hetero-barriers \cite{Coldrenbook}, Auger recombination \cite{Braithwaite1995, Coldrenbook} and inter-valence band absorption \cite{Childs1986, Adams1999} (Fig. \ref{fig4}a), all three of which exponentially increase with temperature. Of the three carrier loss mechanisms noted, Auger recombination and inter-valence band recombination both decrease exponentially with the material's band gap \cite{Adams1999, Coldrenbook}. Hence, shorter wavelength lasers ($<$1 \textmu m) are affected far less by these nonradiative loss processes than longer wavelength lasers ($>$1.2 \textmu m), which are currently used in most telecom applications. In addition, the material systems grown on GaAs substrates used for near-IR to visible wavelength lasers have a favorably larger conduction band offset than that of the InP system of longer wavelength, resulting in higher quantum well barriers and therefore better carrier confinement at elevated temperature \cite{Coldrenbook}. Due to the combination of the above effects, the shorter wavelength photonic platform is expected to display superior high temperature performance (Fig. \ref{fig4}b), which could significantly reduce power consumption by operating with only passive cooling.

To study thermal performance, our heterogeneous III-V/SiN FP lasers were characterized by LI measurements at stage temperatures from 25$^{\circ}$C up to 185$^{\circ}$C, as shown in Fig. \ref{fig4}c. Continuous-wave lasing was achieved up to 185$^{\circ}$C, which is the highest operation temperature among all lasers integrated on a silicon chip to date, significantly higher than the previous record (150$^{\circ}$C)  \cite{Jones2019}. Threshold currents up to 90$^{\circ}$C are well described by an exponential model with a characteristic T$_0$ of 148 K (Fig. \ref{fig4}d), which is on par with the best thermal performance among lasers on native substrate \cite{Derry1992}. At 85$^\circ$C, power at 80 mA and slope efficiency were measured as 12.5 mW and 18.1 \% compared to 14.5 mW and 16.7 \% at room temperature, suggesting only 14 \% degradation in maximum power and an 8 \% improvement in slope efficiency. Additionally, the lasing threshold current remains below 20 mA at 80$^{\circ}$C. 

Such high temperature performance also provides an ultrawide frequency tuning range for FP lasers without relying on any external photonic structure. Spectral measurements indicate significant red-shifting of the lasing wavelength window at a rate of 0.33 nm/K, with a maximum lasing wavelength of 1044.5 nm at 185 C, thus covering more than 50 nm, as shown in Fig. \ref{fig4}e-f.

Beyond simply lasing, the III-V/SiN heterogeneous platform also for the first time demonstrates integrated narrow linewidth lasers at elevated temperature, which greatly benefits many applications including coherent communications in datacenters, remote sensing or metrology in harsh environments. Ring resonator-based tunable lasers (similar to those discussed in the previous section) were characterized. Phase noise measurements were carried out at temperatures from 35$^{\circ}$C up to 145$^{\circ}$C (see Methods). The best overall fundamental linewidth measured was $<$7 kHz, and a linewidth of $<$10 kHz was measured at 145$^{\circ}$C. Only minimal linewidth degradation is observed (see Fig. \ref{fig4}e). It is worth mentioning that integrating both III-V and SiN on the same substrate, ensures a robust coupling between the gain and the external cavity regardless of the temperature. Other linewidth narrowing methods, such as hybrid integration with chip-to-chip butt-coupling \cite{Jin2021, Li2021}, suffer from positional misalignment between different substrates due to thermal expansion mismatch over such a wide temperature range.  \newline

\noindent
\textbf{Discussion} \\
\\
In this work, we have demonstrated a novel photonics platform with profound significance for any field which stands to gain from manipulating short wavelength light. By achieving direct III-V to SiN coupling, the wavelength range of silicon photonics can be extended down to green wavelengths with the gain materials that can be directly grown on GaAs substrate (GaP, InGaP, AlGaAs), and down into blue, violet and UV range by incorporating GaN-based material. With the ultra-low-loss SiN waveguide we have recently demonstrated \cite{Morin2021}, it will be possible to produce high performance PICs at these wavelengths as well. By employing high-$Q$ SiN cavities, fully integrated nonlinear systems may also be realized on this platform, such as microcomb sources \cite{shen2020, Gaeta2019}, stimulated Brillouin lasers \cite{Gundavarapu2018}, and strong frequency conversion systems \cite{Lu2020}. Due to the flexibility of this integration strategy, other materials, such as LiNbO$_3$, AlN, SiC and AlGaAs, can be used intermittently as the media for passive waveguides, which can further enrich the toolbox of integrated photonics. 

This new family of fully integrated PICs, with broad wavelength coverage, versatile functionalities and high performance have the potential to re-write the map of photonics applications. In atomic physics, short wavelength PICs will support on-chip atomic clocks and quantum computing with trapped ion qubits \cite{Mehta2020}. With a platform spanning the vast wavelength range from visible to telecom, coherent links can be designed to support octave spanning self-reference systems for time-frequency metrology \cite{Kues2019} and visible-telecom entanglement in quantum communication \cite{Reimer2016}. In the consumer market, this technology will also create many new opportunities. The exceptional high temperature performance of this platform will relax the cooling requirements of photonic devices, providing an energy-efficient solution for data-centers and photonic computation. By combining highly coherent light sources at visible with low loss optical phase arrays \cite{Poulton2016}, the III-V/SiN heterogeneous photonics platform can potentially remove the bulky lens imaging systems from AR/VR equipment, making it lighter and more power-efficient. 

Finally, since the fabrication of this platform is compatible with existing photonic foundries producing heterogeneous III-V/Si photonics, we expect that this technology will soon be adopted for larger scale high-volume production. As the material cost of SiN-on-Insulator (SiNOI) is lower than that of Silicon-on-Insulator (SOI), this development will make III-V/SiN economically preferable to the now-ubiquitous III-V/Si, reducing costs throughout the industry and truly revolutionizing integrated photonics.\newline

\vspace{6pt}
\noindent \textbf{Methods}\\
\begin{footnotesize}
%\noindent \textbf{Design and fabrication of the devices. } 

\vspace{6pt}
\noindent \textbf{Noise characterization} \\
The delayed self-heterodyne phase noise measurements were performed using a 1 km delay line and a Brimrose TEM-110-10-55-980-2FP AOM on opposing arms of a Mach-Zehnder interferometer. The two output signals were sent to Newport 1801 low noise photoreceivers and recorded with a Tektronix 5 Series mixed signal oscilloscope for cross-correlation analysis as in \cite{Wang2010.09248}. During the measurement, all laser inputs and the thermo-electric cooler of the measurement stage (see below) were controlled by Lightwave ILX LDX-3620B ultra-low noise battery current sources.

\vspace{6pt}
\noindent \textbf{High temperature measurement}\\ 
The high temperature measurement stage consisted of four stages: a heater, a heat spreader, a thermo electric cooler, and the measurements stage. Additionally, polyimid tape was applied to reduce heat-flow to the air, and a hood-like structure of aluminum was added to shield the device under test from air currents. Temperature was monitored using a Vescent SLICE-QTC controller with EPCOS-TDK B57540G1103F005 thermistors in the stage and the heat spreader. Output power was measured with a calibrated Newport 819C-UV-5.3-CAL integrating sphere. Lasing wavelength was determined by measuring spectra with a Yokogawa AQ6374 Optical Spectrum Analyzer.

\vspace{6pt}
\noindent\textbf{Data availability}\\
%The authors declare that the data supporting the findings of this study are available within the article and its Supplementary Information. All raw data are available from the corresponding author upon reasonable request.
The data that supports the plots within this paper and other findings of this study are available from the corresponding authors upon reasonable request. 

\vspace{6pt}
\noindent\textbf{Code availability}\\
The codes that support the findings of this study are available from the corresponding authors upon reasonable request.

\end{footnotesize}
\vspace{20pt}

%_______REFERENCE____________%
\bibliography{shortwavelength.bib}

\begin{thebibliography}{55}%
\makeatletter
\providecommand \@ifxundefined [1]{%
 \@ifx{#1\undefined}
}%
\providecommand \@ifnum [1]{%
 \ifnum #1\expandafter \@firstoftwo
 \else \expandafter \@secondoftwo
 \fi
}%
\providecommand \@ifx [1]{%
 \ifx #1\expandafter \@firstoftwo
 \else \expandafter \@secondoftwo
 \fi
}%
\providecommand \natexlab [1]{#1}%
\providecommand \enquote  [1]{``#1''}%
\providecommand \bibnamefont  [1]{#1}%
\providecommand \bibfnamefont [1]{#1}%
\providecommand \citenamefont [1]{#1}%
\providecommand \href@noop [0]{\@secondoftwo}%
\providecommand \href [0]{\begingroup \@sanitize@url \@href}%
\providecommand \@href[1]{\@@startlink{#1}\@@href}%
\providecommand \@@href[1]{\endgroup#1\@@endlink}%
\providecommand \@sanitize@url [0]{\catcode `\\12\catcode `\$12\catcode
  `\&12\catcode `\#12\catcode `\^12\catcode `\_12\catcode `\%12\relax}%
\providecommand \@@startlink[1]{}%
\providecommand \@@endlink[0]{}%
\providecommand \url  [0]{\begingroup\@sanitize@url \@url }%
\providecommand \@url [1]{\endgroup\@href {#1}{\urlprefix }}%
\providecommand \urlprefix  [0]{URL }%
\providecommand \Eprint [0]{\href }%
\providecommand \doibase [0]{http://dx.doi.org/}%
\providecommand \selectlanguage [0]{\@gobble}%
\providecommand \bibinfo  [0]{\@secondoftwo}%
\providecommand \bibfield  [0]{\@secondoftwo}%
\providecommand \translation [1]{[#1]}%
\providecommand \BibitemOpen [0]{}%
\providecommand \bibitemStop [0]{}%
\providecommand \bibitemNoStop [0]{.\EOS\space}%
\providecommand \EOS [0]{\spacefactor3000\relax}%
\providecommand \BibitemShut  [1]{\csname bibitem#1\endcsname}%
\let\auto@bib@innerbib\@empty
%</preamble>
\bibitem [{\citenamefont {Merz}\ \emph {et~al.}(1985)\citenamefont {Merz},
  \citenamefont {Yuan},\ and\ \citenamefont {Vawter}}]{Merz1985}%
  \BibitemOpen
  \bibfield  {author} {\bibinfo {author} {\bibfnamefont {J.~L.}\ \bibnamefont
  {Merz}}, \bibinfo {author} {\bibfnamefont {Y.~R.}\ \bibnamefont {Yuan}}, \
  and\ \bibinfo {author} {\bibfnamefont {G.~A.}\ \bibnamefont {Vawter}},\
  }\href {\doibase 10.1117/12.7973457} {\bibfield  {journal} {\bibinfo
  {journal} {Optical Engineering}\ }\textbf {\bibinfo {volume} {24}},\ \bibinfo
  {pages} {242214} (\bibinfo {year} {1985})}\BibitemShut {NoStop}%
\bibitem [{\citenamefont {Komljenovic}\ \emph {et~al.}(2018)\citenamefont
  {Komljenovic}, \citenamefont {Huang}, \citenamefont {Pintus}, \citenamefont
  {Tran}, \citenamefont {Davenport},\ and\ \citenamefont
  {Bowers}}]{Komljenovic2018}%
  \BibitemOpen
  \bibfield  {author} {\bibinfo {author} {\bibfnamefont {T.}~\bibnamefont
  {Komljenovic}}, \bibinfo {author} {\bibfnamefont {D.}~\bibnamefont {Huang}},
  \bibinfo {author} {\bibfnamefont {P.}~\bibnamefont {Pintus}}, \bibinfo
  {author} {\bibfnamefont {M.~A.}\ \bibnamefont {Tran}}, \bibinfo {author}
  {\bibfnamefont {M.~L.}\ \bibnamefont {Davenport}}, \ and\ \bibinfo {author}
  {\bibfnamefont {J.~E.}\ \bibnamefont {Bowers}},\ }\href {\doibase
  10.1109/jproc.2018.2864668} {\bibfield  {journal} {\bibinfo  {journal}
  {Proceedings of the IEEE}\ }\textbf {\bibinfo {volume} {106}},\ \bibinfo
  {pages} {2246} (\bibinfo {year} {2018})}\BibitemShut {NoStop}%
\bibitem [{\citenamefont {Margalit}\ \emph {et~al.}(2021)\citenamefont
  {Margalit}, \citenamefont {Xiang}, \citenamefont {Bowers}, \citenamefont
  {Bjorlin}, \citenamefont {Blum},\ and\ \citenamefont
  {Bowers}}]{Margalit2021}%
  \BibitemOpen
  \bibfield  {author} {\bibinfo {author} {\bibfnamefont {N.}~\bibnamefont
  {Margalit}}, \bibinfo {author} {\bibfnamefont {C.}~\bibnamefont {Xiang}},
  \bibinfo {author} {\bibfnamefont {S.~M.}\ \bibnamefont {Bowers}}, \bibinfo
  {author} {\bibfnamefont {A.}~\bibnamefont {Bjorlin}}, \bibinfo {author}
  {\bibfnamefont {R.}~\bibnamefont {Blum}}, \ and\ \bibinfo {author}
  {\bibfnamefont {J.~E.}\ \bibnamefont {Bowers}},\ }\href {\doibase
  10.1063/5.0050117} {\bibfield  {journal} {\bibinfo  {journal} {Appl. Phys.
  Lett.}\ }\textbf {\bibinfo {volume} {118}},\ \bibinfo {pages} {220501}
  (\bibinfo {year} {2021})}\BibitemShut {NoStop}%
\bibitem [{\citenamefont {Zhang}\ \emph {et~al.}(2021)\citenamefont {Zhang},
  \citenamefont {Xing},\ and\ \citenamefont {Huang}}]{Zhang2021}%
  \BibitemOpen
  \bibfield  {author} {\bibinfo {author} {\bibfnamefont {Q.}~\bibnamefont
  {Zhang}}, \bibinfo {author} {\bibfnamefont {Z.}~\bibnamefont {Xing}}, \ and\
  \bibinfo {author} {\bibfnamefont {D.}~\bibnamefont {Huang}},\ }\href
  {\doibase 10.3390/photonics8090363} {\bibfield  {journal} {\bibinfo
  {journal} {Photonics}\ }\textbf {\bibinfo {volume} {8}},\ \bibinfo {pages}
  {363} (\bibinfo {year} {2021})}\BibitemShut {NoStop}%
\bibitem [{\citenamefont {Poulton}\ \emph {et~al.}(2017)\citenamefont
  {Poulton}, \citenamefont {Yaacobi}, \citenamefont {Cole}, \citenamefont
  {Byrd}, \citenamefont {Raval}, \citenamefont {Vermeulen},\ and\ \citenamefont
  {Watts}}]{Poulton2017}%
  \BibitemOpen
  \bibfield  {author} {\bibinfo {author} {\bibfnamefont {C.~V.}\ \bibnamefont
  {Poulton}}, \bibinfo {author} {\bibfnamefont {A.}~\bibnamefont {Yaacobi}},
  \bibinfo {author} {\bibfnamefont {D.~B.}\ \bibnamefont {Cole}}, \bibinfo
  {author} {\bibfnamefont {M.~J.}\ \bibnamefont {Byrd}}, \bibinfo {author}
  {\bibfnamefont {M.}~\bibnamefont {Raval}}, \bibinfo {author} {\bibfnamefont
  {D.}~\bibnamefont {Vermeulen}}, \ and\ \bibinfo {author} {\bibfnamefont
  {M.~R.}\ \bibnamefont {Watts}},\ }\href {\doibase 10.1364/ol.42.004091}
  {\bibfield  {journal} {\bibinfo  {journal} {Optics Letters}\ }\textbf
  {\bibinfo {volume} {42}},\ \bibinfo {pages} {4091} (\bibinfo {year}
  {2017})}\BibitemShut {NoStop}%
\bibitem [{\citenamefont {Qiang}\ \emph {et~al.}(2018)\citenamefont {Qiang},
  \citenamefont {Zhou}, \citenamefont {Wang}, \citenamefont {Wilkes},
  \citenamefont {Loke}, \citenamefont {O'Gara}, \citenamefont {Kling},
  \citenamefont {Marshall}, \citenamefont {Santagati}, \citenamefont {Ralph},
  \citenamefont {Wang}, \citenamefont {O'Brien}, \citenamefont {Thompson},\
  and\ \citenamefont {Matthews}}]{Qiang2018}%
  \BibitemOpen
  \bibfield  {author} {\bibinfo {author} {\bibfnamefont {X.}~\bibnamefont
  {Qiang}}, \bibinfo {author} {\bibfnamefont {X.}~\bibnamefont {Zhou}},
  \bibinfo {author} {\bibfnamefont {J.}~\bibnamefont {Wang}}, \bibinfo {author}
  {\bibfnamefont {C.~M.}\ \bibnamefont {Wilkes}}, \bibinfo {author}
  {\bibfnamefont {T.}~\bibnamefont {Loke}}, \bibinfo {author} {\bibfnamefont
  {S.}~\bibnamefont {O'Gara}}, \bibinfo {author} {\bibfnamefont
  {L.}~\bibnamefont {Kling}}, \bibinfo {author} {\bibfnamefont {G.~D.}\
  \bibnamefont {Marshall}}, \bibinfo {author} {\bibfnamefont {R.}~\bibnamefont
  {Santagati}}, \bibinfo {author} {\bibfnamefont {T.~C.}\ \bibnamefont
  {Ralph}}, \bibinfo {author} {\bibfnamefont {J.~B.}\ \bibnamefont {Wang}},
  \bibinfo {author} {\bibfnamefont {J.~L.}\ \bibnamefont {O'Brien}}, \bibinfo
  {author} {\bibfnamefont {M.~G.}\ \bibnamefont {Thompson}}, \ and\ \bibinfo
  {author} {\bibfnamefont {J.~C.~F.}\ \bibnamefont {Matthews}},\ }\href
  {\doibase 10.1038/s41566-018-0236-y} {\bibfield  {journal} {\bibinfo
  {journal} {Nature Photonics}\ }\textbf {\bibinfo {volume} {12}},\ \bibinfo
  {pages} {534} (\bibinfo {year} {2018})}\BibitemShut {NoStop}%
\bibitem [{\citenamefont {D'Agostino}\ \emph {et~al.}(2015)\citenamefont
  {D'Agostino}, \citenamefont {Carnicella}, \citenamefont {Ciminelli},
  \citenamefont {Thijs}, \citenamefont {Veldhoven}, \citenamefont {Ambrosius},\
  and\ \citenamefont {Smit}}]{Agostino2015}%
  \BibitemOpen
  \bibfield  {author} {\bibinfo {author} {\bibfnamefont {D.}~\bibnamefont
  {D'Agostino}}, \bibinfo {author} {\bibfnamefont {G.}~\bibnamefont
  {Carnicella}}, \bibinfo {author} {\bibfnamefont {C.}~\bibnamefont
  {Ciminelli}}, \bibinfo {author} {\bibfnamefont {P.}~\bibnamefont {Thijs}},
  \bibinfo {author} {\bibfnamefont {P.~J.}\ \bibnamefont {Veldhoven}}, \bibinfo
  {author} {\bibfnamefont {H.}~\bibnamefont {Ambrosius}}, \ and\ \bibinfo
  {author} {\bibfnamefont {M.}~\bibnamefont {Smit}},\ }\href {\doibase
  10.1364/OE.23.025143} {\bibfield  {journal} {\bibinfo  {journal} {Optics
  Express}\ }\textbf {\bibinfo {volume} {23}},\ \bibinfo {pages} {25143}
  (\bibinfo {year} {2015})}\BibitemShut {NoStop}%
\bibitem [{\citenamefont {Ferguson}\ \emph {et~al.}(2006)\citenamefont
  {Ferguson}, \citenamefont {Kuver}, \citenamefont {Heaton}, \citenamefont
  {Zhou}, \citenamefont {Snowden},\ and\ \citenamefont
  {Iezekiel}}]{Ferguson2006}%
  \BibitemOpen
  \bibfield  {author} {\bibinfo {author} {\bibfnamefont {A.}~\bibnamefont
  {Ferguson}}, \bibinfo {author} {\bibfnamefont {A.}~\bibnamefont {Kuver}},
  \bibinfo {author} {\bibfnamefont {J.}~\bibnamefont {Heaton}}, \bibinfo
  {author} {\bibfnamefont {Y.}~\bibnamefont {Zhou}}, \bibinfo {author}
  {\bibfnamefont {C.}~\bibnamefont {Snowden}}, \ and\ \bibinfo {author}
  {\bibfnamefont {S.}~\bibnamefont {Iezekiel}},\ }\href@noop {} {\bibfield
  {journal} {\bibinfo  {journal} {IEE Proceedings-Optoelectronics}\ }\textbf
  {\bibinfo {volume} {153}},\ \bibinfo {pages} {51} (\bibinfo {year}
  {2006})}\BibitemShut {NoStop}%
\bibitem [{\citenamefont {Biberman}\ \emph {et~al.}(2012)\citenamefont
  {Biberman}, \citenamefont {Shaw}, \citenamefont {Timurdogan}, \citenamefont
  {Wright},\ and\ \citenamefont {Watts}}]{Biberman2012}%
  \BibitemOpen
  \bibfield  {author} {\bibinfo {author} {\bibfnamefont {A.}~\bibnamefont
  {Biberman}}, \bibinfo {author} {\bibfnamefont {M.~J.}\ \bibnamefont {Shaw}},
  \bibinfo {author} {\bibfnamefont {E.}~\bibnamefont {Timurdogan}}, \bibinfo
  {author} {\bibfnamefont {J.~B.}\ \bibnamefont {Wright}}, \ and\ \bibinfo
  {author} {\bibfnamefont {M.~R.}\ \bibnamefont {Watts}},\ }\href {\doibase
  10.1364/ol.37.004236} {\bibfield  {journal} {\bibinfo  {journal} {Optics
  Letters}\ }\textbf {\bibinfo {volume} {37}},\ \bibinfo {pages} {4236}
  (\bibinfo {year} {2012})}\BibitemShut {NoStop}%
\bibitem [{\citenamefont {Bellegarde}\ \emph {et~al.}(2018)\citenamefont
  {Bellegarde}, \citenamefont {Erwine~Pargon}, \citenamefont {Petit-Etienne},
  \citenamefont {Hugues}, \citenamefont {Robin-Brosse}, \citenamefont
  {Hartmann},\ and\ \citenamefont {Lyan}}]{Bellegarde2018}%
  \BibitemOpen
  \bibfield  {author} {\bibinfo {author} {\bibfnamefont {C.}~\bibnamefont
  {Bellegarde}}, \bibinfo {author} {\bibfnamefont {C.~S.}\ \bibnamefont
  {Erwine~Pargon}}, \bibinfo {author} {\bibfnamefont {C.}~\bibnamefont
  {Petit-Etienne}}, \bibinfo {author} {\bibfnamefont {V.}~\bibnamefont
  {Hugues}}, \bibinfo {author} {\bibfnamefont {D.}~\bibnamefont
  {Robin-Brosse}}, \bibinfo {author} {\bibfnamefont {J.-M.}\ \bibnamefont
  {Hartmann}}, \ and\ \bibinfo {author} {\bibfnamefont {P.}~\bibnamefont
  {Lyan}},\ }\href@noop {} {\bibfield  {journal} {\bibinfo  {journal} {IEEE
  Photonics Technology Letters}\ }\textbf {\bibinfo {volume} {30}},\ \bibinfo
  {pages} {591} (\bibinfo {year} {2018})}\BibitemShut {NoStop}%
\bibitem [{\citenamefont {Chauhan}\ \emph {et~al.}(2021)\citenamefont
  {Chauhan}, \citenamefont {Isichenko}, \citenamefont {Liu}, \citenamefont
  {Wang}, \citenamefont {Zhao}, \citenamefont {Behunin}, \citenamefont
  {Rakich}, \citenamefont {Jayich}, \citenamefont {Fertig}, \citenamefont
  {Hoyt},\ and\ \citenamefont {Blumenthal}}]{Chauhan2021}%
  \BibitemOpen
  \bibfield  {author} {\bibinfo {author} {\bibfnamefont {N.}~\bibnamefont
  {Chauhan}}, \bibinfo {author} {\bibfnamefont {A.}~\bibnamefont {Isichenko}},
  \bibinfo {author} {\bibfnamefont {K.}~\bibnamefont {Liu}}, \bibinfo {author}
  {\bibfnamefont {J.}~\bibnamefont {Wang}}, \bibinfo {author} {\bibfnamefont
  {Q.}~\bibnamefont {Zhao}}, \bibinfo {author} {\bibfnamefont {R.~O.}\
  \bibnamefont {Behunin}}, \bibinfo {author} {\bibfnamefont {P.~T.}\
  \bibnamefont {Rakich}}, \bibinfo {author} {\bibfnamefont {A.~M.}\
  \bibnamefont {Jayich}}, \bibinfo {author} {\bibfnamefont {C.}~\bibnamefont
  {Fertig}}, \bibinfo {author} {\bibfnamefont {C.~W.}\ \bibnamefont {Hoyt}}, \
  and\ \bibinfo {author} {\bibfnamefont {D.~J.}\ \bibnamefont {Blumenthal}},\
  }\href {\doibase 10.1038/s41467-021-24926-8} {\bibfield  {journal} {\bibinfo
  {journal} {Nature Communications}\ }\textbf {\bibinfo {volume} {12}},\
  \bibinfo {pages} {1} (\bibinfo {year} {2021})}\BibitemShut {NoStop}%
\bibitem [{\citenamefont {Morin}\ \emph {et~al.}(2021)\citenamefont {Morin},
  \citenamefont {Chang}, \citenamefont {Jin}, \citenamefont {Li}, \citenamefont
  {Guo}, \citenamefont {Park}, \citenamefont {Tran}, \citenamefont
  {Komljenovic},\ and\ \citenamefont {Bowers}}]{Morin2021}%
  \BibitemOpen
  \bibfield  {author} {\bibinfo {author} {\bibfnamefont {T.~J.}\ \bibnamefont
  {Morin}}, \bibinfo {author} {\bibfnamefont {L.}~\bibnamefont {Chang}},
  \bibinfo {author} {\bibfnamefont {W.}~\bibnamefont {Jin}}, \bibinfo {author}
  {\bibfnamefont {C.}~\bibnamefont {Li}}, \bibinfo {author} {\bibfnamefont
  {J.}~\bibnamefont {Guo}}, \bibinfo {author} {\bibfnamefont {H.}~\bibnamefont
  {Park}}, \bibinfo {author} {\bibfnamefont {M.~A.}\ \bibnamefont {Tran}},
  \bibinfo {author} {\bibfnamefont {T.}~\bibnamefont {Komljenovic}}, \ and\
  \bibinfo {author} {\bibfnamefont {J.~E.}\ \bibnamefont {Bowers}},\ }\href
  {\doibase 10.1364/optica.426065} {\bibfield  {journal} {\bibinfo  {journal}
  {Optica}\ }\textbf {\bibinfo {volume} {8}},\ \bibinfo {pages} {755} (\bibinfo
  {year} {2021})}\BibitemShut {NoStop}%
\bibitem [{\citenamefont {Puckett}\ \emph {et~al.}(2021)\citenamefont
  {Puckett}, \citenamefont {Liu}, \citenamefont {Chauhan}, \citenamefont
  {Zhao}, \citenamefont {Jin}, \citenamefont {Cheng}, \citenamefont {Wu},
  \citenamefont {Behunin}, \citenamefont {Rakich}, \citenamefont {Nelson},\
  and\ \citenamefont {Blumenthal}}]{Puckett2021}%
  \BibitemOpen
  \bibfield  {author} {\bibinfo {author} {\bibfnamefont {M.~W.}\ \bibnamefont
  {Puckett}}, \bibinfo {author} {\bibfnamefont {K.}~\bibnamefont {Liu}},
  \bibinfo {author} {\bibfnamefont {N.}~\bibnamefont {Chauhan}}, \bibinfo
  {author} {\bibfnamefont {Q.}~\bibnamefont {Zhao}}, \bibinfo {author}
  {\bibfnamefont {N.}~\bibnamefont {Jin}}, \bibinfo {author} {\bibfnamefont
  {H.}~\bibnamefont {Cheng}}, \bibinfo {author} {\bibfnamefont
  {J.}~\bibnamefont {Wu}}, \bibinfo {author} {\bibfnamefont {R.~O.}\
  \bibnamefont {Behunin}}, \bibinfo {author} {\bibfnamefont {P.~T.}\
  \bibnamefont {Rakich}}, \bibinfo {author} {\bibfnamefont {K.~D.}\
  \bibnamefont {Nelson}}, \ and\ \bibinfo {author} {\bibfnamefont {D.~J.}\
  \bibnamefont {Blumenthal}},\ }\href {\doibase 10.1038/s41467-021-21205-4}
  {\bibfield  {journal} {\bibinfo  {journal} {Nature Communications}\ }\textbf
  {\bibinfo {volume} {12}},\ \bibinfo {pages} {1} (\bibinfo {year}
  {2021})}\BibitemShut {NoStop}%
\bibitem [{\citenamefont {Mehta}\ \emph {et~al.}(2020)\citenamefont {Mehta},
  \citenamefont {Zhang}, \citenamefont {Malinowski}, \citenamefont {Nguyen},
  \citenamefont {Stadler},\ and\ \citenamefont {Home}}]{Mehta2020}%
  \BibitemOpen
  \bibfield  {author} {\bibinfo {author} {\bibfnamefont {K.~K.}\ \bibnamefont
  {Mehta}}, \bibinfo {author} {\bibfnamefont {C.}~\bibnamefont {Zhang}},
  \bibinfo {author} {\bibfnamefont {M.}~\bibnamefont {Malinowski}}, \bibinfo
  {author} {\bibfnamefont {T.-L.}\ \bibnamefont {Nguyen}}, \bibinfo {author}
  {\bibfnamefont {M.}~\bibnamefont {Stadler}}, \ and\ \bibinfo {author}
  {\bibfnamefont {J.~P.}\ \bibnamefont {Home}},\ }\href {\doibase
  10.1038/s41586-020-2823-6} {\bibfield  {journal} {\bibinfo  {journal}
  {Nature}\ }\textbf {\bibinfo {volume} {586}},\ \bibinfo {pages} {533}
  (\bibinfo {year} {2020})}\BibitemShut {NoStop}%
\bibitem [{\citenamefont {Niffenegger}\ \emph {et~al.}(2020)\citenamefont
  {Niffenegger}, \citenamefont {Stuart}, \citenamefont {Sorace-Agaskar},
  \citenamefont {Kharas}, \citenamefont {Bramhavar}, \citenamefont {Bruzewicz},
  \citenamefont {Loh}, \citenamefont {Maxson}, \citenamefont {McConnell},
  \citenamefont {Reens}, \citenamefont {West}, \citenamefont {Sage},\ and\
  \citenamefont {Chiaverini}}]{Niffenegger2020}%
  \BibitemOpen
  \bibfield  {author} {\bibinfo {author} {\bibfnamefont {R.~J.}\ \bibnamefont
  {Niffenegger}}, \bibinfo {author} {\bibfnamefont {J.}~\bibnamefont {Stuart}},
  \bibinfo {author} {\bibfnamefont {C.}~\bibnamefont {Sorace-Agaskar}},
  \bibinfo {author} {\bibfnamefont {D.}~\bibnamefont {Kharas}}, \bibinfo
  {author} {\bibfnamefont {S.}~\bibnamefont {Bramhavar}}, \bibinfo {author}
  {\bibfnamefont {C.~D.}\ \bibnamefont {Bruzewicz}}, \bibinfo {author}
  {\bibfnamefont {W.}~\bibnamefont {Loh}}, \bibinfo {author} {\bibfnamefont
  {R.~T.}\ \bibnamefont {Maxson}}, \bibinfo {author} {\bibfnamefont
  {R.}~\bibnamefont {McConnell}}, \bibinfo {author} {\bibfnamefont
  {D.}~\bibnamefont {Reens}}, \bibinfo {author} {\bibfnamefont {G.~N.}\
  \bibnamefont {West}}, \bibinfo {author} {\bibfnamefont {J.~M.}\ \bibnamefont
  {Sage}}, \ and\ \bibinfo {author} {\bibfnamefont {J.}~\bibnamefont
  {Chiaverini}},\ }\href {\doibase 10.1038/s41586-020-2811-x} {\bibfield
  {journal} {\bibinfo  {journal} {Nature}\ }\textbf {\bibinfo {volume} {586}},\
  \bibinfo {pages} {538} (\bibinfo {year} {2020})}\BibitemShut {NoStop}%
\bibitem [{\citenamefont {Hummon}\ \emph {et~al.}(2018)\citenamefont {Hummon},
  \citenamefont {Kang}, \citenamefont {Bopp}, \citenamefont {Li}, \citenamefont
  {Westly}, \citenamefont {Kim}, \citenamefont {Fredrick}, \citenamefont
  {Diddams}, \citenamefont {Srinivasan}, \citenamefont {Aksyuk},\ and\
  \citenamefont {Kitching}}]{Hummon2018}%
  \BibitemOpen
  \bibfield  {author} {\bibinfo {author} {\bibfnamefont {M.~T.}\ \bibnamefont
  {Hummon}}, \bibinfo {author} {\bibfnamefont {S.}~\bibnamefont {Kang}},
  \bibinfo {author} {\bibfnamefont {D.~G.}\ \bibnamefont {Bopp}}, \bibinfo
  {author} {\bibfnamefont {Q.}~\bibnamefont {Li}}, \bibinfo {author}
  {\bibfnamefont {D.~A.}\ \bibnamefont {Westly}}, \bibinfo {author}
  {\bibfnamefont {S.}~\bibnamefont {Kim}}, \bibinfo {author} {\bibfnamefont
  {C.}~\bibnamefont {Fredrick}}, \bibinfo {author} {\bibfnamefont {S.~A.}\
  \bibnamefont {Diddams}}, \bibinfo {author} {\bibfnamefont {K.}~\bibnamefont
  {Srinivasan}}, \bibinfo {author} {\bibfnamefont {V.}~\bibnamefont {Aksyuk}},
  \ and\ \bibinfo {author} {\bibfnamefont {J.~E.}\ \bibnamefont {Kitching}},\
  }\href {\doibase 10.1364/optica.5.000443} {\bibfield  {journal} {\bibinfo
  {journal} {Optica}\ }\textbf {\bibinfo {volume} {5}},\ \bibinfo {pages} {443}
  (\bibinfo {year} {2018})}\BibitemShut {NoStop}%
\bibitem [{\citenamefont {Masood}\ and\ \citenamefont
  {Egger}(2021)}]{Masood2021}%
  \BibitemOpen
  \bibfield  {author} {\bibinfo {author} {\bibfnamefont {T.}~\bibnamefont
  {Masood}}\ and\ \bibinfo {author} {\bibfnamefont {J.}~\bibnamefont {Egger}},\
  }\href {\doibase 10.1109/jstqe.2021.3093721} {\bibfield  {journal} {\bibinfo
  {journal} {IEEE Journal of Quantum Electronics}\ }\textbf {\bibinfo {volume}
  {27}},\ \bibinfo {pages} {1} (\bibinfo {year} {2021})}\BibitemShut {NoStop}%
\bibitem [{\citenamefont {Zinoviev}\ \emph {et~al.}(2011)\citenamefont
  {Zinoviev}, \citenamefont {Gonzalez-Guerrero}, \citenamefont {Dominguez},\
  and\ \citenamefont {Lechuga}}]{Zinoviev2011}%
  \BibitemOpen
  \bibfield  {author} {\bibinfo {author} {\bibfnamefont {K.~E.}\ \bibnamefont
  {Zinoviev}}, \bibinfo {author} {\bibfnamefont {A.~B.}\ \bibnamefont
  {Gonzalez-Guerrero}}, \bibinfo {author} {\bibfnamefont {C.}~\bibnamefont
  {Dominguez}}, \ and\ \bibinfo {author} {\bibfnamefont {L.~M.}\ \bibnamefont
  {Lechuga}},\ }\href {\doibase 10.1109/jlt.2011.2150734} {\bibfield  {journal}
  {\bibinfo  {journal} {Journal of Lightwave Technology}\ }\textbf {\bibinfo
  {volume} {29}},\ \bibinfo {pages} {1926} (\bibinfo {year}
  {2011})}\BibitemShut {NoStop}%
\bibitem [{\citenamefont {Orieux}\ and\ \citenamefont
  {Diamanti}(2016)}]{Orieux2016}%
  \BibitemOpen
  \bibfield  {author} {\bibinfo {author} {\bibfnamefont {A.}~\bibnamefont
  {Orieux}}\ and\ \bibinfo {author} {\bibfnamefont {E.}~\bibnamefont
  {Diamanti}},\ }\href {\doibase 10.1088/2040-8978/18/8/083002} {\bibfield
  {journal} {\bibinfo  {journal} {Journal of Optics}\ }\textbf {\bibinfo
  {volume} {18}},\ \bibinfo {pages} {083002} (\bibinfo {year}
  {2016})}\BibitemShut {NoStop}%
\bibitem [{\citenamefont {Blumenthal}(2020)}]{Blumenthal2020}%
  \BibitemOpen
  \bibfield  {author} {\bibinfo {author} {\bibfnamefont {D.~J.}\ \bibnamefont
  {Blumenthal}},\ }\href {\doibase 10.1063/1.5131683} {\bibfield  {journal}
  {\bibinfo  {journal} {{APL} Photonics}\ }\textbf {\bibinfo {volume} {5}},\
  \bibinfo {pages} {020903} (\bibinfo {year} {2020})}\BibitemShut {NoStop}%
\bibitem [{\citenamefont {Kr\"{u}ckel}\ \emph {et~al.}(2017)\citenamefont
  {Kr\"{u}ckel}, \citenamefont {F\"{u}l\"{o}p}, \citenamefont {Ye},
  \citenamefont {Andrekson},\ and\ \citenamefont
  {Torres-Company}}]{Kruckel2017}%
  \BibitemOpen
  \bibfield  {author} {\bibinfo {author} {\bibfnamefont {C.~J.}\ \bibnamefont
  {Kr\"{u}ckel}}, \bibinfo {author} {\bibfnamefont {A.}~\bibnamefont
  {F\"{u}l\"{o}p}}, \bibinfo {author} {\bibfnamefont {Z.}~\bibnamefont {Ye}},
  \bibinfo {author} {\bibfnamefont {P.~A.}\ \bibnamefont {Andrekson}}, \ and\
  \bibinfo {author} {\bibfnamefont {V.}~\bibnamefont {Torres-Company}},\ }\href
  {\doibase 10.1364/oe.25.015370} {\bibfield  {journal} {\bibinfo  {journal}
  {Optics Express}\ }\textbf {\bibinfo {volume} {25}},\ \bibinfo {pages}
  {15370} (\bibinfo {year} {2017})}\BibitemShut {NoStop}%
\bibitem [{\citenamefont {Jin}\ \emph {et~al.}(2021)\citenamefont {Jin},
  \citenamefont {Yang}, \citenamefont {Chang}, \citenamefont {Shen},
  \citenamefont {Wang}, \citenamefont {Leal}, \citenamefont {Wu}, \citenamefont
  {Gao}, \citenamefont {Feshali}, \citenamefont {Paniccia}, \citenamefont
  {Vahala},\ and\ \citenamefont {Bowers}}]{Jin2021}%
  \BibitemOpen
  \bibfield  {author} {\bibinfo {author} {\bibfnamefont {W.}~\bibnamefont
  {Jin}}, \bibinfo {author} {\bibfnamefont {Q.-F.}\ \bibnamefont {Yang}},
  \bibinfo {author} {\bibfnamefont {L.}~\bibnamefont {Chang}}, \bibinfo
  {author} {\bibfnamefont {B.}~\bibnamefont {Shen}}, \bibinfo {author}
  {\bibfnamefont {H.}~\bibnamefont {Wang}}, \bibinfo {author} {\bibfnamefont
  {M.~A.}\ \bibnamefont {Leal}}, \bibinfo {author} {\bibfnamefont
  {L.}~\bibnamefont {Wu}}, \bibinfo {author} {\bibfnamefont {M.}~\bibnamefont
  {Gao}}, \bibinfo {author} {\bibfnamefont {A.}~\bibnamefont {Feshali}},
  \bibinfo {author} {\bibfnamefont {M.}~\bibnamefont {Paniccia}}, \bibinfo
  {author} {\bibfnamefont {K.~J.}\ \bibnamefont {Vahala}}, \ and\ \bibinfo
  {author} {\bibfnamefont {J.~E.}\ \bibnamefont {Bowers}},\ }\href {\doibase
  10.1038/s41566-021-00761-7} {\bibfield  {journal} {\bibinfo  {journal}
  {Nature Photonics}\ } (\bibinfo {year} {2021}),\
  10.1038/s41566-021-00761-7}\BibitemShut {NoStop}%
\bibitem [{\citenamefont {Xiang}\ \emph {et~al.}(2020)\citenamefont {Xiang},
  \citenamefont {Jin}, \citenamefont {Guo}, \citenamefont {Peters},
  \citenamefont {Kennedy}, \citenamefont {Selvidge}, \citenamefont {Morton},\
  and\ \citenamefont {Bowers}}]{Xiang2020}%
  \BibitemOpen
  \bibfield  {author} {\bibinfo {author} {\bibfnamefont {C.}~\bibnamefont
  {Xiang}}, \bibinfo {author} {\bibfnamefont {W.}~\bibnamefont {Jin}}, \bibinfo
  {author} {\bibfnamefont {J.}~\bibnamefont {Guo}}, \bibinfo {author}
  {\bibfnamefont {J.~D.}\ \bibnamefont {Peters}}, \bibinfo {author}
  {\bibfnamefont {M.~J.}\ \bibnamefont {Kennedy}}, \bibinfo {author}
  {\bibfnamefont {J.}~\bibnamefont {Selvidge}}, \bibinfo {author}
  {\bibfnamefont {P.~A.}\ \bibnamefont {Morton}}, \ and\ \bibinfo {author}
  {\bibfnamefont {J.~E.}\ \bibnamefont {Bowers}},\ }\href {\doibase
  10.1364/optica.384026} {\bibfield  {journal} {\bibinfo  {journal} {Optica}\
  }\textbf {\bibinfo {volume} {7}},\ \bibinfo {pages} {20} (\bibinfo {year}
  {2020})}\BibitemShut {NoStop}%
\bibitem [{\citenamefont {de~Beeck}\ \emph {et~al.}(2020)\citenamefont
  {de~Beeck}, \citenamefont {Haq}, \citenamefont {Elsinger}, \citenamefont
  {Gocalinska}, \citenamefont {Pelucchi}, \citenamefont {Corbett},
  \citenamefont {Roelkens},\ and\ \citenamefont {Kuyken}}]{Beeck2020}%
  \BibitemOpen
  \bibfield  {author} {\bibinfo {author} {\bibfnamefont {C.~O.}\ \bibnamefont
  {de~Beeck}}, \bibinfo {author} {\bibfnamefont {B.}~\bibnamefont {Haq}},
  \bibinfo {author} {\bibfnamefont {L.}~\bibnamefont {Elsinger}}, \bibinfo
  {author} {\bibfnamefont {A.}~\bibnamefont {Gocalinska}}, \bibinfo {author}
  {\bibfnamefont {E.}~\bibnamefont {Pelucchi}}, \bibinfo {author}
  {\bibfnamefont {B.}~\bibnamefont {Corbett}}, \bibinfo {author} {\bibfnamefont
  {G.}~\bibnamefont {Roelkens}}, \ and\ \bibinfo {author} {\bibfnamefont
  {B.}~\bibnamefont {Kuyken}},\ }\href@noop {} {\bibfield  {journal} {\bibinfo
  {journal} {Optica}\ }\textbf {\bibinfo {volume} {7}},\ \bibinfo {pages} {386}
  (\bibinfo {year} {2020})}\BibitemShut {NoStop}%
\bibitem [{\citenamefont {Fang}\ \emph {et~al.}(2006)\citenamefont {Fang},
  \citenamefont {Park}, \citenamefont {Cohen}, \citenamefont {Jones},
  \citenamefont {Paniccia},\ and\ \citenamefont {Bowers}}]{Fang2006}%
  \BibitemOpen
  \bibfield  {author} {\bibinfo {author} {\bibfnamefont {A.~W.}\ \bibnamefont
  {Fang}}, \bibinfo {author} {\bibfnamefont {H.}~\bibnamefont {Park}}, \bibinfo
  {author} {\bibfnamefont {O.}~\bibnamefont {Cohen}}, \bibinfo {author}
  {\bibfnamefont {R.}~\bibnamefont {Jones}}, \bibinfo {author} {\bibfnamefont
  {M.~J.}\ \bibnamefont {Paniccia}}, \ and\ \bibinfo {author} {\bibfnamefont
  {J.~E.}\ \bibnamefont {Bowers}},\ }\href {\doibase 10.1364/oe.14.009203}
  {\bibfield  {journal} {\bibinfo  {journal} {Optics Express}\ }\textbf
  {\bibinfo {volume} {14}},\ \bibinfo {pages} {9203} (\bibinfo {year}
  {2006})}\BibitemShut {NoStop}%
\bibitem [{\citenamefont {Park}(2020)}]{park2020pat}%
  \BibitemOpen
  \bibfield  {author} {\bibinfo {author} {\bibfnamefont {H.}~\bibnamefont
  {Park}},\ }\href@noop {} {\enquote {\bibinfo {title} {Integrated active
  devices with improved optical coupling to dielectric waveguides},}\ }
  (\bibinfo {year} {2020}),\ \bibinfo {note} {uS Patent 10,718,898}\BibitemShut
  {NoStop}%
\bibitem [{\citenamefont {Park}\ \emph {et~al.}(2020)\citenamefont {Park},
  \citenamefont {Zhang}, \citenamefont {Tran},\ and\ \citenamefont
  {Komljenovic}}]{Park2020}%
  \BibitemOpen
  \bibfield  {author} {\bibinfo {author} {\bibfnamefont {H.}~\bibnamefont
  {Park}}, \bibinfo {author} {\bibfnamefont {C.}~\bibnamefont {Zhang}},
  \bibinfo {author} {\bibfnamefont {M.~A.}\ \bibnamefont {Tran}}, \ and\
  \bibinfo {author} {\bibfnamefont {T.}~\bibnamefont {Komljenovic}},\ }\href
  {\doibase 10.1364/optica.391809} {\bibfield  {journal} {\bibinfo  {journal}
  {Optica}\ }\textbf {\bibinfo {volume} {7}},\ \bibinfo {pages} {336} (\bibinfo
  {year} {2020})}\BibitemShut {NoStop}%
\bibitem [{\citenamefont {Chen}(1970)}]{chen1970}%
  \BibitemOpen
  \bibfield  {author} {\bibinfo {author} {\bibfnamefont {F.-S.}\ \bibnamefont
  {Chen}},\ }\href@noop {} {\bibfield  {journal} {\bibinfo  {journal}
  {Proceedings of the IEEE}\ }\textbf {\bibinfo {volume} {58}},\ \bibinfo
  {pages} {1440} (\bibinfo {year} {1970})}\BibitemShut {NoStop}%
\bibitem [{\citenamefont {Drever}\ \emph {et~al.}(1983)\citenamefont {Drever},
  \citenamefont {Hall}, \citenamefont {Kowalski}, \citenamefont {Hough},
  \citenamefont {Ford}, \citenamefont {Munley},\ and\ \citenamefont
  {Ward}}]{Drever1983}%
  \BibitemOpen
  \bibfield  {author} {\bibinfo {author} {\bibfnamefont {R.}~\bibnamefont
  {Drever}}, \bibinfo {author} {\bibfnamefont {J.~L.}\ \bibnamefont {Hall}},
  \bibinfo {author} {\bibfnamefont {F.}~\bibnamefont {Kowalski}}, \bibinfo
  {author} {\bibfnamefont {J.}~\bibnamefont {Hough}}, \bibinfo {author}
  {\bibfnamefont {G.}~\bibnamefont {Ford}}, \bibinfo {author} {\bibfnamefont
  {A.}~\bibnamefont {Munley}}, \ and\ \bibinfo {author} {\bibfnamefont
  {H.}~\bibnamefont {Ward}},\ }\href@noop {} {\bibfield  {journal} {\bibinfo
  {journal} {Applied Physics B}\ }\textbf {\bibinfo {volume} {31}},\ \bibinfo
  {pages} {97} (\bibinfo {year} {1983})}\BibitemShut {NoStop}%
\bibitem [{\citenamefont {Metcalf}\ and\ \citenamefont {van~der
  Straten}(1999)}]{LaserCoolingbook}%
  \BibitemOpen
  \bibfield  {author} {\bibinfo {author} {\bibfnamefont {H.~J.}\ \bibnamefont
  {Metcalf}}\ and\ \bibinfo {author} {\bibfnamefont {P.}~\bibnamefont {van~der
  Straten}},\ }\href@noop {} {\emph {\bibinfo {title} {Laser Cooling and
  Trapping}}}\ (\bibinfo  {publisher} {Springer},\ \bibinfo {year}
  {1999})\BibitemShut {NoStop}%
\bibitem [{\citenamefont {Tran}\ \emph {et~al.}(2019)\citenamefont {Tran},
  \citenamefont {Huang},\ and\ \citenamefont {Bowers}}]{Tran2019}%
  \BibitemOpen
  \bibfield  {author} {\bibinfo {author} {\bibfnamefont {M.~A.}\ \bibnamefont
  {Tran}}, \bibinfo {author} {\bibfnamefont {D.}~\bibnamefont {Huang}}, \ and\
  \bibinfo {author} {\bibfnamefont {J.~E.}\ \bibnamefont {Bowers}},\ }\href
  {\doibase 10.1063/1.5124254} {\bibfield  {journal} {\bibinfo  {journal}
  {{APL} Photonics}\ }\textbf {\bibinfo {volume} {4}},\ \bibinfo {pages}
  {111101} (\bibinfo {year} {2019})}\BibitemShut {NoStop}%
\bibitem [{\citenamefont {Wieman}\ and\ \citenamefont
  {Hollberg}(1991)}]{Wieman1991}%
  \BibitemOpen
  \bibfield  {author} {\bibinfo {author} {\bibfnamefont {C.~E.}\ \bibnamefont
  {Wieman}}\ and\ \bibinfo {author} {\bibfnamefont {L.}~\bibnamefont
  {Hollberg}},\ }\href {\doibase 10.1063/1.1142305} {\bibfield  {journal}
  {\bibinfo  {journal} {Contemporary Physics}\ }\textbf {\bibinfo {volume}
  {62}},\ \bibinfo {pages} {1} (\bibinfo {year} {1991})}\BibitemShut {NoStop}%
\bibitem [{\citenamefont {Arnold}\ \emph {et~al.}(1998)\citenamefont {Arnold},
  \citenamefont {Wilson},\ and\ \citenamefont {Boshier}}]{Arnold1998}%
  \BibitemOpen
  \bibfield  {author} {\bibinfo {author} {\bibfnamefont {A.~S.}\ \bibnamefont
  {Arnold}}, \bibinfo {author} {\bibfnamefont {J.~S.}\ \bibnamefont {Wilson}},
  \ and\ \bibinfo {author} {\bibfnamefont {M.~G.}\ \bibnamefont {Boshier}},\
  }\href {\doibase 10.1063/1.1148756} {\bibfield  {journal} {\bibinfo
  {journal} {Review of Scientific Instruments}\ }\textbf {\bibinfo {volume}
  {69}},\ \bibinfo {pages} {1236} (\bibinfo {year} {1998})}\BibitemShut
  {NoStop}%
\bibitem [{\citenamefont {Liu}\ and\ \citenamefont
  {Littman}(1981)}]{Littman1981}%
  \BibitemOpen
  \bibfield  {author} {\bibinfo {author} {\bibfnamefont {K.}~\bibnamefont
  {Liu}}\ and\ \bibinfo {author} {\bibfnamefont {M.~G.}\ \bibnamefont
  {Littman}},\ }\href {\doibase 10.1364/ol.6.000117} {\bibfield  {journal}
  {\bibinfo  {journal} {Optics Letters}\ }\textbf {\bibinfo {volume} {6}},\
  \bibinfo {pages} {117} (\bibinfo {year} {1981})}\BibitemShut {NoStop}%
\bibitem [{\citenamefont {Vogel}\ \emph {et~al.}(1999)\citenamefont {Vogel},
  \citenamefont {Dinneen}, \citenamefont {Gallagher},\ and\ \citenamefont
  {Hall}}]{Vogel1999}%
  \BibitemOpen
  \bibfield  {author} {\bibinfo {author} {\bibfnamefont {K.}~\bibnamefont
  {Vogel}}, \bibinfo {author} {\bibfnamefont {T.}~\bibnamefont {Dinneen}},
  \bibinfo {author} {\bibfnamefont {A.}~\bibnamefont {Gallagher}}, \ and\
  \bibinfo {author} {\bibfnamefont {J.}~\bibnamefont {Hall}},\ }\href {\doibase
  10.1109/19.769671} {\bibfield  {journal} {\bibinfo  {journal} {IEEE Trans.
  Instrum. Meas}\ }\textbf {\bibinfo {volume} {48}},\ \bibinfo {pages} {618}
  (\bibinfo {year} {1999})}\BibitemShut {NoStop}%
\bibitem [{\citenamefont {McFerran}\ and\ \citenamefont
  {Luiten}(2010)}]{McFerran2010}%
  \BibitemOpen
  \bibfield  {author} {\bibinfo {author} {\bibfnamefont {J.~J.}\ \bibnamefont
  {McFerran}}\ and\ \bibinfo {author} {\bibfnamefont {A.~N.}\ \bibnamefont
  {Luiten}},\ }\href {\doibase 10.1364/JOSAB.27.000277} {\bibfield  {journal}
  {\bibinfo  {journal} {J. Opt. Soc. Am. B}\ }\textbf {\bibinfo {volume}
  {27}},\ \bibinfo {pages} {277} (\bibinfo {year} {2010})}\BibitemShut
  {NoStop}%
\bibitem [{\citenamefont {Uttam}\ and\ \citenamefont
  {Culshaw}(1985)}]{uttam_FMCW1985}%
  \BibitemOpen
  \bibfield  {author} {\bibinfo {author} {\bibfnamefont {D.}~\bibnamefont
  {Uttam}}\ and\ \bibinfo {author} {\bibfnamefont {B.}~\bibnamefont
  {Culshaw}},\ }\href@noop {} {\bibfield  {journal} {\bibinfo  {journal}
  {Journal of Lightwave Technology}\ }\textbf {\bibinfo {volume} {3}},\
  \bibinfo {pages} {971} (\bibinfo {year} {1985})}\BibitemShut {NoStop}%
\bibitem [{\citenamefont {van Rees}\ \emph {et~al.}(2020)\citenamefont {van
  Rees}, \citenamefont {Fan}, \citenamefont {Geskus}, \citenamefont {Klein},
  \citenamefont {Oldenbeuving}, \citenamefont {van~der Slot},\ and\
  \citenamefont {Boller}}]{Rees2020}%
  \BibitemOpen
  \bibfield  {author} {\bibinfo {author} {\bibfnamefont {A.}~\bibnamefont {van
  Rees}}, \bibinfo {author} {\bibfnamefont {Y.}~\bibnamefont {Fan}}, \bibinfo
  {author} {\bibfnamefont {D.}~\bibnamefont {Geskus}}, \bibinfo {author}
  {\bibfnamefont {E.~J.}\ \bibnamefont {Klein}}, \bibinfo {author}
  {\bibfnamefont {R.~M.}\ \bibnamefont {Oldenbeuving}}, \bibinfo {author}
  {\bibfnamefont {P.~J.~M.}\ \bibnamefont {van~der Slot}}, \ and\ \bibinfo
  {author} {\bibfnamefont {K.-J.}\ \bibnamefont {Boller}},\ }\href {\doibase
  10.1364/oe.386356} {\bibfield  {journal} {\bibinfo  {journal} {Optics
  Express}\ }\textbf {\bibinfo {volume} {28}},\ \bibinfo {pages} {5669}
  (\bibinfo {year} {2020})}\BibitemShut {NoStop}%
\bibitem [{\citenamefont {Jones}\ \emph {et~al.}(2019)\citenamefont {Jones},
  \citenamefont {Doussiere}, \citenamefont {Driscoll}, \citenamefont {Lin},
  \citenamefont {Yu}, \citenamefont {Akulova}, \citenamefont {Komljenovic},\
  and\ \citenamefont {Bowers}}]{Jones2019}%
  \BibitemOpen
  \bibfield  {author} {\bibinfo {author} {\bibfnamefont {R.}~\bibnamefont
  {Jones}}, \bibinfo {author} {\bibfnamefont {P.}~\bibnamefont {Doussiere}},
  \bibinfo {author} {\bibfnamefont {J.~B.}\ \bibnamefont {Driscoll}}, \bibinfo
  {author} {\bibfnamefont {W.}~\bibnamefont {Lin}}, \bibinfo {author}
  {\bibfnamefont {H.}~\bibnamefont {Yu}}, \bibinfo {author} {\bibfnamefont
  {Y.}~\bibnamefont {Akulova}}, \bibinfo {author} {\bibfnamefont
  {T.}~\bibnamefont {Komljenovic}}, \ and\ \bibinfo {author} {\bibfnamefont
  {J.~E.}\ \bibnamefont {Bowers}},\ }\href {\doibase
  10.1109/mnano.2019.2891369} {\bibfield  {journal} {\bibinfo  {journal} {IEEE
  Nanotechnology Magazine}\ }\textbf {\bibinfo {volume} {13}},\ \bibinfo
  {pages} {17} (\bibinfo {year} {2019})}\BibitemShut {NoStop}%
\bibitem [{\citenamefont {Jones}(2018)}]{Jones2018}%
  \BibitemOpen
  \bibfield  {author} {\bibinfo {author} {\bibfnamefont {N.}~\bibnamefont
  {Jones}},\ }\href {\doibase 10.1038/d41586-018-06610-y} {\bibfield  {journal}
  {\bibinfo  {journal} {Nature}\ }\textbf {\bibinfo {volume} {561}},\ \bibinfo
  {pages} {163} (\bibinfo {year} {2018})}\BibitemShut {NoStop}%
\bibitem [{\citenamefont {Piprek}\ \emph {et~al.}(2000)\citenamefont {Piprek},
  \citenamefont {Abraham},\ and\ \citenamefont {Bowers}}]{Piprek2000}%
  \BibitemOpen
  \bibfield  {author} {\bibinfo {author} {\bibfnamefont {J.}~\bibnamefont
  {Piprek}}, \bibinfo {author} {\bibfnamefont {P.}~\bibnamefont {Abraham}}, \
  and\ \bibinfo {author} {\bibfnamefont {J.}~\bibnamefont {Bowers}},\ }\href
  {\doibase 10.1109/3.825885} {\bibfield  {journal} {\bibinfo  {journal} {IEEE
  Journal of Quantum Electronics}\ }\textbf {\bibinfo {volume} {36}},\ \bibinfo
  {pages} {366} (\bibinfo {year} {2000})}\BibitemShut {NoStop}%
\bibitem [{\citenamefont {Coldren}\ \emph {et~al.}(2012)\citenamefont
  {Coldren}, \citenamefont {Corzine},\ and\ \citenamefont
  {Mashanovitch}}]{Coldrenbook}%
  \BibitemOpen
  \bibfield  {author} {\bibinfo {author} {\bibfnamefont {L.~A.}\ \bibnamefont
  {Coldren}}, \bibinfo {author} {\bibfnamefont {S.~W.}\ \bibnamefont
  {Corzine}}, \ and\ \bibinfo {author} {\bibfnamefont {M.~L.}\ \bibnamefont
  {Mashanovitch}},\ }\href@noop {} {\emph {\bibinfo {title} {Diode Lasers and
  Photonic Integrated Circuits}}}\ (\bibinfo  {publisher} {Wiley},\ \bibinfo
  {year} {2012})\BibitemShut {NoStop}%
\bibitem [{\citenamefont {Braithwaite}\ \emph {et~al.}(1995)\citenamefont
  {Braithwaite}, \citenamefont {Silver}, \citenamefont {Wilkinson},
  \citenamefont {O'Reilly},\ and\ \citenamefont {Adams}}]{Braithwaite1995}%
  \BibitemOpen
  \bibfield  {author} {\bibinfo {author} {\bibfnamefont {J.}~\bibnamefont
  {Braithwaite}}, \bibinfo {author} {\bibfnamefont {M.}~\bibnamefont {Silver}},
  \bibinfo {author} {\bibfnamefont {V.~A.}\ \bibnamefont {Wilkinson}}, \bibinfo
  {author} {\bibfnamefont {E.~P.}\ \bibnamefont {O'Reilly}}, \ and\ \bibinfo
  {author} {\bibfnamefont {A.~R.}\ \bibnamefont {Adams}},\ }\href {\doibase
  10.1063/1.114916} {\bibfield  {journal} {\bibinfo  {journal} {Applied Physics
  Letters}\ }\textbf {\bibinfo {volume} {67}},\ \bibinfo {pages} {3546}
  (\bibinfo {year} {1995})}\BibitemShut {NoStop}%
\bibitem [{\citenamefont {Childs}\ \emph {et~al.}(1986)\citenamefont {Childs},
  \citenamefont {Brand},\ and\ \citenamefont {Abram}}]{Childs1986}%
  \BibitemOpen
  \bibfield  {author} {\bibinfo {author} {\bibfnamefont {G.~N.}\ \bibnamefont
  {Childs}}, \bibinfo {author} {\bibfnamefont {S.}~\bibnamefont {Brand}}, \
  and\ \bibinfo {author} {\bibfnamefont {R.~A.}\ \bibnamefont {Abram}},\ }\href
  {\doibase 10.1088/0268-1242/1/2/004} {\bibfield  {journal} {\bibinfo
  {journal} {Semiconductor Science and Technology}\ }\textbf {\bibinfo {volume}
  {1}},\ \bibinfo {pages} {116} (\bibinfo {year} {1986})}\BibitemShut {NoStop}%
\bibitem [{\citenamefont {Adams}\ \emph {et~al.}(1999)\citenamefont {Adams},
  \citenamefont {O'Reilly},\ and\ \citenamefont {Silver}}]{Adams1999}%
  \BibitemOpen
  \bibfield  {author} {\bibinfo {author} {\bibfnamefont {A.~R.}\ \bibnamefont
  {Adams}}, \bibinfo {author} {\bibfnamefont {E.~P.}\ \bibnamefont {O'Reilly}},
  \ and\ \bibinfo {author} {\bibfnamefont {M.}~\bibnamefont {Silver}}\
  }(\bibinfo  {publisher} {Elsevier},\ \bibinfo {year} {1999})\ pp.\ \bibinfo
  {pages} {123--176}\BibitemShut {NoStop}%
\bibitem [{\citenamefont {Derry}\ \emph {et~al.}(1992)\citenamefont {Derry},
  \citenamefont {Hager}, \citenamefont {Chiu}, \citenamefont {Booher},
  \citenamefont {Miao},\ and\ \citenamefont {Hong}}]{Derry1992}%
  \BibitemOpen
  \bibfield  {author} {\bibinfo {author} {\bibfnamefont {P.}~\bibnamefont
  {Derry}}, \bibinfo {author} {\bibfnamefont {H.}~\bibnamefont {Hager}},
  \bibinfo {author} {\bibfnamefont {K.}~\bibnamefont {Chiu}}, \bibinfo {author}
  {\bibfnamefont {D.}~\bibnamefont {Booher}}, \bibinfo {author} {\bibfnamefont
  {E.}~\bibnamefont {Miao}}, \ and\ \bibinfo {author} {\bibfnamefont
  {C.}~\bibnamefont {Hong}},\ }\href@noop {} {\bibfield  {journal} {\bibinfo
  {journal} {IEEE Photonics Technology Letters}\ }\textbf {\bibinfo {volume}
  {4}},\ \bibinfo {pages} {1189} (\bibinfo {year} {1992})}\BibitemShut
  {NoStop}%
\bibitem [{\citenamefont {Li}\ \emph {et~al.}(2021)\citenamefont {Li},
  \citenamefont {Jin}, \citenamefont {Wu}, \citenamefont {Chang}, \citenamefont
  {Wang}, \citenamefont {Shen}, \citenamefont {Yuan}, \citenamefont {Feshali},
  \citenamefont {Paniccia}, \citenamefont {Vahala} \emph {et~al.}}]{Li2021}%
  \BibitemOpen
  \bibfield  {author} {\bibinfo {author} {\bibfnamefont {B.}~\bibnamefont
  {Li}}, \bibinfo {author} {\bibfnamefont {W.}~\bibnamefont {Jin}}, \bibinfo
  {author} {\bibfnamefont {L.}~\bibnamefont {Wu}}, \bibinfo {author}
  {\bibfnamefont {L.}~\bibnamefont {Chang}}, \bibinfo {author} {\bibfnamefont
  {H.}~\bibnamefont {Wang}}, \bibinfo {author} {\bibfnamefont {B.}~\bibnamefont
  {Shen}}, \bibinfo {author} {\bibfnamefont {Z.}~\bibnamefont {Yuan}}, \bibinfo
  {author} {\bibfnamefont {A.}~\bibnamefont {Feshali}}, \bibinfo {author}
  {\bibfnamefont {M.}~\bibnamefont {Paniccia}}, \bibinfo {author}
  {\bibfnamefont {K.~J.}\ \bibnamefont {Vahala}},  \emph {et~al.},\ }\href@noop
  {} {\bibfield  {journal} {\bibinfo  {journal} {Optics Letters}\ }\textbf
  {\bibinfo {volume} {46}},\ \bibinfo {pages} {5201} (\bibinfo {year}
  {2021})}\BibitemShut {NoStop}%
\bibitem [{\citenamefont {Shen}\ \emph {et~al.}(2020)\citenamefont {Shen},
  \citenamefont {Chang}, \citenamefont {Liu}, \citenamefont {Wang},
  \citenamefont {Qi-Fan~Yang}, \citenamefont {Wang}, \citenamefont {He},
  \citenamefont {Liu}, \citenamefont {Xie}, \citenamefont {Guo}, \citenamefont
  {Kinghorn}, \citenamefont {Wu}, \citenamefont {Ji}, \citenamefont
  {Kippenberg}, \citenamefont {Vahala},\ and\ \citenamefont
  {Bowers}}]{shen2020}%
  \BibitemOpen
  \bibfield  {author} {\bibinfo {author} {\bibfnamefont {B.}~\bibnamefont
  {Shen}}, \bibinfo {author} {\bibfnamefont {L.}~\bibnamefont {Chang}},
  \bibinfo {author} {\bibfnamefont {J.}~\bibnamefont {Liu}}, \bibinfo {author}
  {\bibfnamefont {H.}~\bibnamefont {Wang}}, \bibinfo {author} {\bibfnamefont
  {C.~X.}\ \bibnamefont {Qi-Fan~Yang}}, \bibinfo {author} {\bibfnamefont
  {R.~N.}\ \bibnamefont {Wang}}, \bibinfo {author} {\bibfnamefont
  {J.}~\bibnamefont {He}}, \bibinfo {author} {\bibfnamefont {T.}~\bibnamefont
  {Liu}}, \bibinfo {author} {\bibfnamefont {W.}~\bibnamefont {Xie}}, \bibinfo
  {author} {\bibfnamefont {J.}~\bibnamefont {Guo}}, \bibinfo {author}
  {\bibfnamefont {D.}~\bibnamefont {Kinghorn}}, \bibinfo {author}
  {\bibfnamefont {L.}~\bibnamefont {Wu}}, \bibinfo {author} {\bibfnamefont
  {Q.-X.}\ \bibnamefont {Ji}}, \bibinfo {author} {\bibfnamefont {T.~J.}\
  \bibnamefont {Kippenberg}}, \bibinfo {author} {\bibfnamefont
  {K.}~\bibnamefont {Vahala}}, \ and\ \bibinfo {author} {\bibfnamefont {J.~E.}\
  \bibnamefont {Bowers}},\ }\href {\doibase 10.1038/s41586-020-2358-x}
  {\bibfield  {journal} {\bibinfo  {journal} {Nature}\ }\textbf {\bibinfo
  {volume} {582}},\ \bibinfo {pages} {365} (\bibinfo {year}
  {2020})}\BibitemShut {NoStop}%
\bibitem [{\citenamefont {Gaeta}\ \emph {et~al.}(2019)\citenamefont {Gaeta},
  \citenamefont {Lipson},\ and\ \citenamefont {Kippenberg}}]{Gaeta2019}%
  \BibitemOpen
  \bibfield  {author} {\bibinfo {author} {\bibfnamefont {A.~L.}\ \bibnamefont
  {Gaeta}}, \bibinfo {author} {\bibfnamefont {M.}~\bibnamefont {Lipson}}, \
  and\ \bibinfo {author} {\bibfnamefont {T.~J.}\ \bibnamefont {Kippenberg}},\
  }\href {\doibase 10.1038/s41566-019-0358-x} {\bibfield  {journal} {\bibinfo
  {journal} {Nature Photonics}\ }\textbf {\bibinfo {volume} {13}},\ \bibinfo
  {pages} {158} (\bibinfo {year} {2019})}\BibitemShut {NoStop}%
\bibitem [{\citenamefont {Gundavarapu}\ \emph {et~al.}(2018)\citenamefont
  {Gundavarapu}, \citenamefont {Brodnik}, \citenamefont {Puckett},
  \citenamefont {Huffman}, \citenamefont {Bose}, \citenamefont {Behunin},
  \citenamefont {Wu}, \citenamefont {Qiu}, \citenamefont {Pinho}, \citenamefont
  {Chauhan}, \citenamefont {Nohava}, \citenamefont {Rakich}, \citenamefont
  {Nelson}, \citenamefont {Salit},\ and\ \citenamefont
  {Blumenthal}}]{Gundavarapu2018}%
  \BibitemOpen
  \bibfield  {author} {\bibinfo {author} {\bibfnamefont {S.}~\bibnamefont
  {Gundavarapu}}, \bibinfo {author} {\bibfnamefont {G.~M.}\ \bibnamefont
  {Brodnik}}, \bibinfo {author} {\bibfnamefont {M.}~\bibnamefont {Puckett}},
  \bibinfo {author} {\bibfnamefont {T.}~\bibnamefont {Huffman}}, \bibinfo
  {author} {\bibfnamefont {D.}~\bibnamefont {Bose}}, \bibinfo {author}
  {\bibfnamefont {R.}~\bibnamefont {Behunin}}, \bibinfo {author} {\bibfnamefont
  {J.}~\bibnamefont {Wu}}, \bibinfo {author} {\bibfnamefont {T.}~\bibnamefont
  {Qiu}}, \bibinfo {author} {\bibfnamefont {C.}~\bibnamefont {Pinho}}, \bibinfo
  {author} {\bibfnamefont {N.}~\bibnamefont {Chauhan}}, \bibinfo {author}
  {\bibfnamefont {J.}~\bibnamefont {Nohava}}, \bibinfo {author} {\bibfnamefont
  {P.~T.}\ \bibnamefont {Rakich}}, \bibinfo {author} {\bibfnamefont {K.~D.}\
  \bibnamefont {Nelson}}, \bibinfo {author} {\bibfnamefont {M.}~\bibnamefont
  {Salit}}, \ and\ \bibinfo {author} {\bibfnamefont {D.~J.}\ \bibnamefont
  {Blumenthal}},\ }\href {\doibase 10.1038/s41566-018-0313-2} {\bibfield
  {journal} {\bibinfo  {journal} {Nature Photonics}\ }\textbf {\bibinfo
  {volume} {13}},\ \bibinfo {pages} {60} (\bibinfo {year} {2018})}\BibitemShut
  {NoStop}%
\bibitem [{\citenamefont {Lu}\ \emph {et~al.}(2020)\citenamefont {Lu},
  \citenamefont {Moille}, \citenamefont {Rao}, \citenamefont {Westly},\ and\
  \citenamefont {Srinivasan}}]{Lu2020}%
  \BibitemOpen
  \bibfield  {author} {\bibinfo {author} {\bibfnamefont {X.}~\bibnamefont
  {Lu}}, \bibinfo {author} {\bibfnamefont {G.}~\bibnamefont {Moille}}, \bibinfo
  {author} {\bibfnamefont {A.}~\bibnamefont {Rao}}, \bibinfo {author}
  {\bibfnamefont {D.~A.}\ \bibnamefont {Westly}}, \ and\ \bibinfo {author}
  {\bibfnamefont {K.}~\bibnamefont {Srinivasan}},\ }\href {\doibase
  10.1038/s41566-020-00708-4} {\bibfield  {journal} {\bibinfo  {journal}
  {Nature Photonics}\ }\textbf {\bibinfo {volume} {15}},\ \bibinfo {pages}
  {131} (\bibinfo {year} {2020})}\BibitemShut {NoStop}%
\bibitem [{\citenamefont {Kues}\ \emph {et~al.}(2019)\citenamefont {Kues},
  \citenamefont {Reimer}, \citenamefont {Lukens}, \citenamefont {Munro},
  \citenamefont {Weiner}, \citenamefont {Moss},\ and\ \citenamefont
  {Morandotti}}]{Kues2019}%
  \BibitemOpen
  \bibfield  {author} {\bibinfo {author} {\bibfnamefont {M.}~\bibnamefont
  {Kues}}, \bibinfo {author} {\bibfnamefont {C.}~\bibnamefont {Reimer}},
  \bibinfo {author} {\bibfnamefont {J.~M.}\ \bibnamefont {Lukens}}, \bibinfo
  {author} {\bibfnamefont {W.~J.}\ \bibnamefont {Munro}}, \bibinfo {author}
  {\bibfnamefont {A.~M.}\ \bibnamefont {Weiner}}, \bibinfo {author}
  {\bibfnamefont {D.~J.}\ \bibnamefont {Moss}}, \ and\ \bibinfo {author}
  {\bibfnamefont {R.}~\bibnamefont {Morandotti}},\ }\href {\doibase
  10.1038/s41566-019-0363-0} {\bibfield  {journal} {\bibinfo  {journal} {Nature
  Photonics}\ }\textbf {\bibinfo {volume} {13}},\ \bibinfo {pages} {170}
  (\bibinfo {year} {2019})}\BibitemShut {NoStop}%
\bibitem [{\citenamefont {Reimer}\ \emph {et~al.}(2016)\citenamefont {Reimer},
  \citenamefont {Kues}, \citenamefont {Roztocki}, \citenamefont {Wetzel},
  \citenamefont {Grazioso}, \citenamefont {Little}, \citenamefont {Chu},
  \citenamefont {Johnston}, \citenamefont {Bromberg}, \citenamefont {Caspani},
  \citenamefont {Moss},\ and\ \citenamefont {Morandotti}}]{Reimer2016}%
  \BibitemOpen
  \bibfield  {author} {\bibinfo {author} {\bibfnamefont {C.}~\bibnamefont
  {Reimer}}, \bibinfo {author} {\bibfnamefont {M.}~\bibnamefont {Kues}},
  \bibinfo {author} {\bibfnamefont {P.}~\bibnamefont {Roztocki}}, \bibinfo
  {author} {\bibfnamefont {B.}~\bibnamefont {Wetzel}}, \bibinfo {author}
  {\bibfnamefont {F.}~\bibnamefont {Grazioso}}, \bibinfo {author}
  {\bibfnamefont {B.~E.}\ \bibnamefont {Little}}, \bibinfo {author}
  {\bibfnamefont {S.~T.}\ \bibnamefont {Chu}}, \bibinfo {author} {\bibfnamefont
  {T.}~\bibnamefont {Johnston}}, \bibinfo {author} {\bibfnamefont
  {Y.}~\bibnamefont {Bromberg}}, \bibinfo {author} {\bibfnamefont
  {L.}~\bibnamefont {Caspani}}, \bibinfo {author} {\bibfnamefont {D.~J.}\
  \bibnamefont {Moss}}, \ and\ \bibinfo {author} {\bibfnamefont
  {R.}~\bibnamefont {Morandotti}},\ }\href {\doibase 10.1126/science.aad8532}
  {\bibfield  {journal} {\bibinfo  {journal} {Science}\ }\textbf {\bibinfo
  {volume} {351}},\ \bibinfo {pages} {1176} (\bibinfo {year}
  {2016})}\BibitemShut {NoStop}%
\bibitem [{\citenamefont {Poulton}\ \emph {et~al.}(2016)\citenamefont
  {Poulton}, \citenamefont {Byrd}, \citenamefont {Raval}, \citenamefont {Su},
  \citenamefont {Li}, \citenamefont {Timurdogan}, \citenamefont {Coolbaugh},
  \citenamefont {Vermeulen},\ and\ \citenamefont {Watts}}]{Poulton2016}%
  \BibitemOpen
  \bibfield  {author} {\bibinfo {author} {\bibfnamefont {C.~V.}\ \bibnamefont
  {Poulton}}, \bibinfo {author} {\bibfnamefont {M.~J.}\ \bibnamefont {Byrd}},
  \bibinfo {author} {\bibfnamefont {M.}~\bibnamefont {Raval}}, \bibinfo
  {author} {\bibfnamefont {Z.}~\bibnamefont {Su}}, \bibinfo {author}
  {\bibfnamefont {N.}~\bibnamefont {Li}}, \bibinfo {author} {\bibfnamefont
  {E.}~\bibnamefont {Timurdogan}}, \bibinfo {author} {\bibfnamefont
  {D.}~\bibnamefont {Coolbaugh}}, \bibinfo {author} {\bibfnamefont
  {D.}~\bibnamefont {Vermeulen}}, \ and\ \bibinfo {author} {\bibfnamefont
  {M.~R.}\ \bibnamefont {Watts}},\ }\href {\doibase 10.1364/ol.42.000021}
  {\bibfield  {journal} {\bibinfo  {journal} {Optics Letters}\ }\textbf
  {\bibinfo {volume} {42}},\ \bibinfo {pages} {21} (\bibinfo {year}
  {2016})}\BibitemShut {NoStop}%
\bibitem [{\citenamefont {Wang}\ \emph {et~al.}(2020)\citenamefont {Wang},
  \citenamefont {Wu}, \citenamefont {Yuan},\ and\ \citenamefont
  {Vahala}}]{Wang2010.09248}%
  \BibitemOpen
  \bibfield  {author} {\bibinfo {author} {\bibfnamefont {H.}~\bibnamefont
  {Wang}}, \bibinfo {author} {\bibfnamefont {L.}~\bibnamefont {Wu}}, \bibinfo
  {author} {\bibfnamefont {Z.}~\bibnamefont {Yuan}}, \ and\ \bibinfo {author}
  {\bibfnamefont {K.}~\bibnamefont {Vahala}},\ }\href@noop {} {\enquote
  {\bibinfo {title} {Towards milli-hertz laser frequency noise on a chip},}\ }
  (\bibinfo {year} {2020}),\ \Eprint {http://arxiv.org/abs/arXiv:2010.09248}
  {arXiv:2010.09248} \BibitemShut {NoStop}%
\end{thebibliography}%
% \bibliographystyle{naturemag}
% \bibliographystyle{naturesaa}

%%_______REFERENCE_______%%

%--------------------------------------------------------------------------
\vspace{12pt}
\begin{footnotesize}

\vspace{6pt}
\noindent \textbf{Acknowledgment}

\noindent 
The authors thank David Kinghorn for assistance with measurements, Sonya Palmer for fruitful discussion and Lillian McKinney, Brian Long and Yujun Chen for graphic sketches. We are also grateful to Larry Coldren for discussion of high temperature laser performance as well as David Weld and Jianwei Wang for discussion of atomic physics applications. A portion of this work was performed in the UCSB Nanofabrication Facility, an open access laboratory. Part of this work and material (related to UCSB and Caltech) is based upon work supported by the Defense Advanced Research Projects Agency (DARPA) under Contract No. HR001-20-2-0044.
Any opinions, findings and conclusions or recommendations expressed in this material are those of the author(s) and do not necessarily reflect the views of the Defense Advanced Research Projects Agency (DARPA).\\

%\\vspace{6pt}
%\\noindent \textbf{Author contributions}

%\All devices were designed and fabricated by Nexus (M.T., C.Z., S.B., W.L., Z.Z., T.K., H.P.). High temperature measurements and linewidth measurements were performed by UCSB (T.M, L.C., J.G., J.B.). Phase noise theory and experimental design were performed by Caltech (Z.Y., H.W., B.S., L.W., K.V.).

% Joel Guo tuned the lasers with his magical hands.

%\vspace{6pt}
\noindent \textbf{Author contributions}
\noindent

\noindent All devices were designed by H.P., T.K., C.Z. and M.T., and fabricated by C.Z., M.T., W.L. and G.K. Device characterization was performed by H.P, S.B., A.M. and Z.Z. High temperature and noise characterization were performed by T.M., M.T., L.C. and J.G., with the assistance of Z.Y., H.W., B.S. and L.W. Theoretical phase noise investigation was conducted by Z.Y., H.W., B.S. and L.W. The manuscript was prepared by M.T., T.M. and L.C., with the assistance from all of the other authors. T.K. supervised all operations at Nexus Photonics including design, fabrication and characterizations, and L.C., J.B. and K.V. supervised advanced characterization at UCSB and Caltech.
\noindent

\vspace{6pt}
\noindent
\textbf{Additional information} 

\noindent Supplementary information is available in the online version of the paper. Reprints and permissions information is available online. Correspondence and requests for materials should be addressed to the corresponding authors.

\vspace{6pt}
\noindent \textbf{Competing interests} 

\noindent J.B. is a cofounder of Nexus Photonics.
\end{footnotesize}
%--------------------------------------------------------------------------

\end{document}